%% file: main.tex
\documentclass[12pt,a4paper]{article}
\usepackage[utf8]{inputenc}
\usepackage[dvips]{graphics}
\RequirePackage{lineno} 
\usepackage[pdftex]{color, graphicx}
\usepackage{amsmath, amsthm}
\usepackage{multirow}
\usepackage{subfigure}
\usepackage{xspace}
\usepackage{cite}
\usepackage{ifthen}
\usepackage{authblk}
\usepackage{lineno}

\RequirePackage{lineno}
\usepackage[pdfauthor={AM+FPGA Team},bookmarks=true]{hyperref} 

\title{Charged Particle Tracking in Real-Time Using a Full-Mesh Data Delivery Architecture and Associative Memory Techniques}

\author[a]{Sudha Ajuha}
\author[a]{Ailton Akira Shinoda}
\author[a]{Lucas Arruda Ramalho}
\author[b]{Guillaume Baulieu}
\author[b]{Gaelle Boudoul}
\author[c]{Massimo Casarsa}
\author[a]{Andre Cascadan}
\author[d]{Emyr Clement}
\author[a]{Thiago Costa de Paiva}
\author[e]{Souvik Das}
\author[f]{Suchandra Dutta}
\author[g]{Ricardo Eusebi}
\author[h]{Giacomo Fedi}
\author[a]{Vitor Finotti Ferreira}
\author[i]{Kristian Hahn}
\author[j]{Zhen Hu}
\author[j*]{Sergo Jindariani}
\author[e]{Jacobo Konigsberg}
\author[j]{Tiehui Liu}
\author[e]{Jia Fu Low}
\author[k]{Emily MacDonald}
\author[j]{Jamieson Olsen}
\author[h]{Fabrizio Palla}
\author[l]{Nicola Pozzobon}
\author[g]{Denis Rathjens}
\author[j]{Luciano Ristori}
\author[l]{Roberto Rossin}
\author[i]{Kevin Sung}
\author[j]{Nhan Tran}
\author[i]{Marco Trovato}
\author[k]{Keith Ulmer}
\author[a]{Mario Vaz}
\author[b]{Sebastien Viret}
\author[j]{Jin-Yuan Wu}
\author[m]{Zijun Xu}
\author[i]{Silvia Zorzetti}

\affil[a]{UNESP - Sao Paulo State University, Sao Paulo, Brazil}
\affil[b]{Institut de Physique Nucleaire de Lyon (IPNL), Lyon, France}
\affil[c]{INFN Sezione di Trieste, Trieste, Italy}
\affil[d]{University of Bristol, Bristol, United Kingdom}
\affil[e]{University of Florida, Gainesville, Florida, USA}
\affil[f]{Saha Institute of Nuclear Physics, HBNI, Kolkata, India}
\affil[g]{Texas A\&M University, College Station, Texas, USA}
\affil[h]{INFN Sezione di Pisa, Pisa, Italy}
\affil[i]{Northwestern University, Evanston, Illinois, USA}
\affil[j]{Fermi National Accelerator Laboratory, Batavia, Illinois, USA}
\affil[k]{University of Colorado Boulder, Boulder, Colorado, USA}
\affil[l]{INFN Sezione di Padova, Universita` di Padova, Padova, Italy}
\affil[m]{Peking University, Beijing, China}

\newcommand{\pt}{\ensuremath{p_{\mathrm{T}}}\xspace}
\newcommand{\Et}{\ensuremath{E_{\mathrm{T}}}\xspace}

\def\antibar#1{\ensuremath{#1\bar{#1}}}%

\def\ttbar{\ensuremath{\mathrm{\antibar{t}}}\xspace }%

\begin{document}
\maketitle
\renewcommand*{\thefootnote}{\fnsymbol{footnote}}
\footnote{Corresponding author, sergo@fnal.gov}

\begin{abstract}
We present a flexible and scalable approach to address the challenges of charged particle track reconstruction in real-time event filters (Level-1 triggers) in collider physics experiments. The method described here is based on a full-mesh architecture for data distribution and relies on the Associative Memory approach to implement a pattern recognition algorithm that quickly identifies and organizes hits associated to trajectories of particles originating from particle collisions. We describe a successful implementation of a demonstration system composed of several innovative hardware and algorithmic elements. The implementation of a full-size system relies on the assumption that an Associative Memory device with the sufficient pattern density becomes available in the future, either through a dedicated ASIC or a modern FPGA. We demonstrate excellent performance in terms of track reconstruction efficiency, purity, momentum resolution, and processing time measured with data from a simulated LHC-like tracking detector.
\end{abstract}  
\tableofcontents

\input{Introduction}
\input{Overview}

\input{Delivery}

\input{PatternRecognition}
\input{Demonstration}

\input{Results}

\input{Summary}

\bibliographystyle{auto_generated}
\bibliography{bibliography}

\appendix
\input{Hardware}
\end{document}

%% file: Introduction.tex
\section{Introduction \label{sec:intro}}
Most high-energy physics experiments, such as those carried out at hadron colliders, are exposed to ever increasing luminosity. Due to the large size, complexity, and high granularity of their various detector components, the amount of data acquired by these experiments can be staggering. As a result, recording and processing every collision event is practically impossible. It is also not necessary since most of the events are not of interest to the physics program. Collider experiments have dealt with this difficulty by implementing real-time event selection methods, called {\em triggers}, that reduce the amount of data to levels manageable for storage and offline processing while recording most of the events relevant for physics analyses. These triggers are critical to the success of the physics program of the experiments, as they allow selecting only events with which important measurements or discoveries can be made. To this end, the real-time reconstruction of all physics objects in each event has to be carried out as fast and as accurately as possible. Triggers are typically designed so that decisions are made in sequential steps. Each successive step, or level, handles a significantly reduced event rate with respect to the previous one and has access to more detailed information.

For example, the trigger systems of the current Large Hadron Collider (LHC) experiments will continue to face increasingly complex requirements. In order to vastly expand the physics reach of the ATLAS and CMS experiments, the LHC and the experimental collaborations are pursuing an immense upgrade program with an expected completion date of 2029. This will usher the so-called HL-LHC era (for High-Luminosity LHC) during which the luminosity will increase by a factor of five relative to the Run~2 instantaneous luminosity, resulting in an average number of proton-proton collisions per bunch crossing, referred to as {\em pileup}, of approximately 200. To contend with such a serious increase in proton-proton collision rates, and the correspondingly high detector occupancy, several components of the ATLAS and CMS experiments will have to be upgraded. In particular, the trigger systems need to undergo a significant redesign~\cite{CERN-LHCC-2017-009,Zabi:2020gjd,CERN-LHCC-2017-020}. Looking even further into the future, the anticipated detector complexity and beam background conditions at the high-energy colliders proposed beyond the LHC (FCC-hh, SppC, muon collider) will be orders of magnitude higher than those at the HL-LHC, further exacerbating the problem of real-time reconstruction.

The ability to reconstruct trajectories of charged particles in the trigger system and to measure their transverse momenta (\pt) with high precision can be a powerful asset as this significantly improves the \pt measurement of leptons and jets, which are otherwise measured much less accurately. This in turn means that the set of collision events with such objects required to pass a certain \pt threshold in the trigger will be selected more accurately, providing control over the trigger selection rate. An alternative to this would be to require higher \pt thresholds, which would lead to a reduction of events relevant to the experiment's physics program. Therefore, the implementation of a track trigger system at HL-LHC and future collider experiments is important to the preservation, and even the expansion, of programs that include extensive explorations at the electroweak scale and the study of new processes with potential for discoveries.

The implementation of a Level-1 (L1) track trigger system is extremely challenging due to the high rate of collision events, the amount of data produced in each event, and the requirement for few microsecond processing time. The expected data rates from the detector to the trigger electronics at the HL-LHC experiments are of the order of 100\,Terabits per second (Tb/s)~\cite{CERN-LHCC-2017-009}. Prior to the start of track reconstruction, the data need to be organized and distributed so that all energy deposits from charged particles, referred to as {\em hits}, from a particular region of the tracker and from the same bunch crossing are located in the same electronics processor board at the same time. To solve this complex problem, a massively parallel data distribution and processing architecture is needed. Such an architecture has to have the capability to simultaneously handle events originating from different bunch crossings (time multiplexing) as well as different physical detector regions within the same bunch crossing (regional multiplexing). It also has to be flexible and scalable in order to accommodate possible future changes, upgrades, or unforeseen operating conditions. The challenge is to reconstruct all tracks above a low \pt threshold (of about 2\,GeV) with high efficiency, purity, and momentum resolution that are as close as possible to the quality of the offline reconstruction.

In what follows, we describe a platform to demonstrate track reconstruction at the L1 trigger for experiments running in an environment similar to what is expected at the HL-LHC. Section~2 provides a general description of the approach and the specific assumptions made for the development of the presented solution. Sections~3 and 4 provide an overview of the full-mesh data delivery scheme and the Associative Memory (AM)+ FPGA approach, and include the studies and novel algorithmic concepts tried in order to optimize the performance of the system. Section~5 describes the hardware and firmware used for the demonstration setup, with its results presented in Section~6. Additional simulation studies to further improve the performance of such a system are presented in Section~7, and the paper concludes with a summary in Section~8. 

The approach presented in this paper is one of three approaches described in the Technical Design Report of the Phase-2 upgrade of the CMS tracker~\cite{CERN-LHCC-2017-009}. However, the focus of this paper is more on the conceptual features of the approach and its scalability, which can serve as guiding principles for addressing L1 tracking needs of future high-energy collider experiments. The studies presented here were performed with a demonstration setup at Fermilab.

%% file: Overview.tex
\section{Overview of the Approach\label{sec:tracker}}
\subsection{Data Processing Stages \label{sec:AMFPGA-Overview}}
The approach described in this paper is directly applicable to a generic tracking detector with silicon sensors; however, it could also be applicable to other pattern recognition problems in which execution speed is paramount.

Our approach utilizes a full-mesh architecture for efficient data distribution and relies on AM, implemented in fast electronics chips~\cite{DellOrso:1988gmu}, to handle pattern recognition and the rapid growth of pattern complexity with hit occupancy. In this paper, we refer to this approach as AM+FPGA; however, it should be noted that the full-mesh architecture is generally compatible with other (non-AM) pattern recognition schemes.

An approach based on AM was implemented successfully in the CDF experiment~\cite{Ashmanskas:2003gf} during Run~2 of the Tevatron in a real-time trigger that selected tracks originating from secondary vertices produced by the decay of bottom and charm hadrons~\cite{CDF:2004jtw, Ristori:2010zz}. The challenge is that real-time track reconstruction at the HL-LHC is much more complex than what the CDF experiment had to contend with: not only is the collision rate much higher, but also the number of electronic channels is orders of magnitude larger. 

The AM+FPGA approach can be described as a sequence of three processing stages: data delivery, pattern recognition, and track fitting. In the data delivery stage, the data from the tracker back-end electronics are formatted and distributed so that all hits that originate in a certain geometric region of the detector and belong to the same bunch crossing, are brought simultaneously to a Data Organizer (DO) unit for common processing. The pattern recognition stage involves selecting those hits that were potentially created by the same charged particle. To do that, each layer of the tracking detector is subdivided into coarse groupings of silicon strips (superstrips) which are then linked across layers into patterns that correspond to probable trajectories of charged particles through the detector. The set of all most likely patterns produced by charged tracks is obtained from simulation and loaded into the Associative Memory, implemented in an ASIC or FPGA. The AM provides a very fast way to simultaneously recognize all those patterns that were produced by the tracks created in the collisions and reject all hits that do not belong to any of those patterns. Each pattern recognized by the Associative Memory is grouped together with all the hits that fall within that pattern to form a {\em road}. The roads are sent to the track fitting stage, where the hits are used to extract track parameters from the coordinates of the stubs~\cite{Ristori:2010zz}. 

Track fitting is implemented in FGPAs~\cite{Clement:2018apn} downstream from the pattern recognition stage. The Combination Builder (CB) generates all possible combinations of hits within each road. The hits from each combination are then input to a number of linearized ${\chi}^2$ track fitters (TF). The coefficients used in the linearized fits are pre-determined from a principal component analysis (PCA) of simulated tracks. Mis-reconstructed tracks that do not correspond to particles are referred to as fake tracks and suppressed by requiring good quality of the fit. Some tracks may be reconstructed multiple times in this procedure. Such duplicate tracks are removed by selecting the track with the best ${\chi}^2$ probability from sets of tracks that contain hits in common.

\subsection{Assumptions \label{sec:ModelAssumptions}}
In this section, we describe the detector configuration and provide key specifications for the demonstration. While it is important to state them here, we stress that the presented approach is largely independent of the specifics of the tracker geometry and of the data formatting schemes in the detector front-end electronics. 

We use a right-handed coordinate system, with the origin at the nominal collision point, the $x$-axis pointing to the center of the collider ring, the $y$-axis pointing up (perpendicular to the collider plane), and the $z$-axis along the counterclockwise beam direction. The polar angle $\theta$ is measured from the positive $z$-axis and the azimuthal angle ($\phi$) is measured from the positive $x$-axis in the $x$-$y$ plane. The radius ($r$) denotes the distance from the $z$-axis and the pseudorapidity $\eta$ is defined as $\eta = -\mathrm{ln}[\mathrm{\tan}(\theta/2)]$.

\begin{figure}
\centering
\includegraphics[width=15cm]{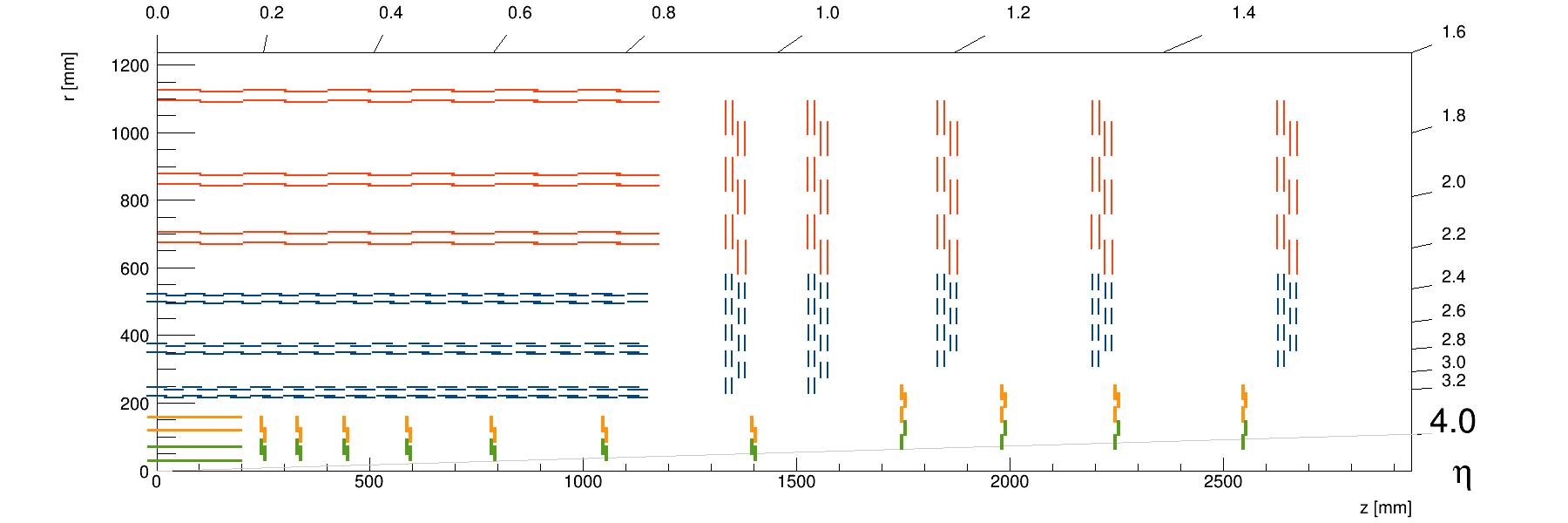}
\caption{One quadrant of the CMS tracker geometry used for the demonstration. Only the Outer Tracker modules shown in red and blue are providing data for L1 track reconstruction.}
\label{fig:tracker}
\end{figure}

For the purpose of the specific study presented here, we consider a geometry consisting of six cylindrical barrel layers in the central region, with modules aligned along the beam direction, and five endcap discs on each side of the barrel, with modules aligned perpendicular to the beam direction. This geometry is shown in Fig.~\ref{fig:tracker} and was initially proposed for the HL-LHC upgrade of the CMS tracker and was later modified~\cite{CERN-LHCC-2017-009}. This paper focuses primarily on track reconstruction in the barrel part of such a detector, corresponding to the pseudorapidity region $|\eta|<1.0$. However the techniques can be easily extended into the endcap region.  

The vast majority of hits in hadron collisions originate from particles with low transverse momentum (from a few MeV to a few GeV). However, for the purpose of L1 lepton and jet reconstruction, it is typically sufficient to reconstruct tracks above a \pt threshold of few GeV. It is therefore desirable that the tracker front-end electronics be capable of rejecting signals from particles below a certain \pt threshold, as this can significantly reduce the number of hits the track trigger has to contend with. The CMS Outer Tracker aims to achieve this by utilizing “\pt modules”~\cite{CERN-LHCC-2017-009}. These modules are composed of two single-sided closely-spaced silicon sensors read out by a set of front-end ASICs. These ASICs correlate the signals from the two sensors of a given tracker module to form the so-called stubs. They also calculate the "stub bend" ($\delta s$), which describes the spacial separation between clusters from the two sensors. They then select stubs with $\delta s$ compatible with particles above a chosen \pt threshold. In this paper, we aim to reconstruct tracks with $\pt>3$~GeV and discuss the implications of reducing this threshold down to $\pt>2$ GeV. Using such thresholds, the hit multiplicity in the tracker is reduced by an order of magnitude, making the pattern recognition problem much more solvable. Similar on-detector data reduction schemes are being proposed for detectors at future colliders. In general, any method that reduces significantly, up front, the amount of data to be contended with by the pattern recognition engine provides a great relief on the resources needed for that stage.

We assume an event input rate of 40~MHz (corresponding to a 25\,ns bunch spacing), and an average of 140 (PU140) or 200 (PU200) inelastic proton-proton collisions per bunch crossing, which roughly corresponds to what is anticipated in the nominal and extended running conditions at the HL-LHC. The effects of data truncation in the detector front-end electronics, when the modules' hit occupancy exceeds the available bandwidth, are considered negligible. The number of stubs read out in each event depends on the physics process, on the assumed pileup amount, and on the stub \pt threshold in the front-end electronics. Each transmitted stub contains its local coordinates and the $\delta s$ value. 

Different simulation samples are generated in order to test the performance of the track trigger. These are implemented in floating point simulation and fixed point emulation. Single particle (muon, pion, and electron) samples are used for generating the AM pattern banks described below and for evaluating the track reconstruction efficiency and track parameter resolution. Samples corresponding to the PU140 and PU200 conditions are used to calculate the reconstructed track rate, as well as the fractions of real, fake, and duplicate tracks in the output sample. Single particles (muons, pions, and electrons) embedded into PU140 and PU200 conditions are studied to assess the impact of pileup on the track reconstruction. Special samples containing top quark pair-production events and multijet events with high transverse momentum jets generated with \textsc{MADGRAPH}~\cite{Alwall:2011uj} are used to study the performance in high-occupancy collision events. Finally,  higher pileup samples with 300 (PU300) and 400 (PU400) interactions per bunch crossing are generated to test the robustness of the system. The \textsc{GEANT4}~\cite{GEANT4:2002zbu} package is used to simulate the response of the CMS detector for all the processes.

In PU200 conditions, on average approximately 10,000 stubs with $\pt>2$~GeV are expected from simulation in each event in the entire tracking detector. The variation of the number of stubs is pretty broad, and the track trigger system should therefore be capable of contending with up to 20,000 stubs from events in the tail of the number of stubs distribution. High energy QCD jets present the most challenging environment for the algorithm. As a note, an isolated single particle (such as an electron or a muon originating from W boson decays) typically adds a negligible number of stubs to an event dominated by a large number of pileup interactions. The presence of a high-energy process such as top quark pair-production increases the stub occupancy by only 10$\%$. However, energetic jets produced in top quark decays populate a localized detector region with very high stub occupancy, and thus can be used to test the track reconstruction efficiency in high stub density environments.

We define the total latency as the time interval between the moment of the bunch crossing and when the last track from that crossing is reconstructed and passed on for downstream processing. As an example, for the CMS HL-LHC track trigger this latency is required to be less than $5\,\mu$s, out of which approximately one $\mu$s is taken by the transmission of  data from the detector to the counting room where the track finder electronics are assumed to be located. This defines a target latency of about $4\,\mu$s for data movement and partitioning as well as for track reconstruction and parameter fitting in the AM+FPGA system.

%% file: Delivery.tex
 \section{Data Delivery \label{sec:DataDelivery}}
As mentioned above, the Level-1 track trigger described here needs to process a new bunch crossing every 25\,ns, corresponding to an input data rate of approximately 100~Tb/s, and reconstruct all tracks with a latency below $4\,\mu$s. The total bandwidth and computing power needed to solve this problem are orders of magnitude larger than previously achieved in a trigger system. Therefore the data distribution and processing architecture needs to be massively parallel.  In such a system, a critical issue is how to quickly and efficiently organize and distribute the data to the right processing unit (engine) for pattern recognition and track fitting. 

\subsection{Trigger Towers}
For the purpose of regional multiplexing, we introduce the concept of {\em trigger towers} as a way of partitioning the entire tracker into $N$ different regions in which track finding can be performed independently. We first define {\em virtual trigger towers} as a subdivision of the track parameter space into $N$ non-overlapping regions. The track parameter space we consider is four-dimensional: $\phi$, $q/\pt$, $\eta$, and $z_{0}$, where $q$ is electric charge of the track. At this level, we consider only tracks that intersect the $z$ axis (have impact parameter of zero) and $z_{0}$ is the $z$ coordinate of the track at $x=0, y=0$. The union of all the $N$ regions covers, with no overlap, the parameter space where we want the track finding efficiency to be optimized.
{\em Physical trigger towers} are then constructed as collections of all the tracker modules that we need to include to obtain full efficiency for all the tracks coming from the parameter space of the corresponding virtual tower.

Adjacent physical towers need to overlap in order to preserve efficiency for tracks crossing their boundary. This means that, to be able to perform track finding in each tower independently, some stubs must be duplicated and sent to multiple towers. A larger overlap means more stubs to be duplicated, larger data traffic and more data processing. The size of the overlap can be minimized by a clever choice of the shape of the virtual trigger towers.

\begin{figure}[h]
\centering
\includegraphics[width=14cm]{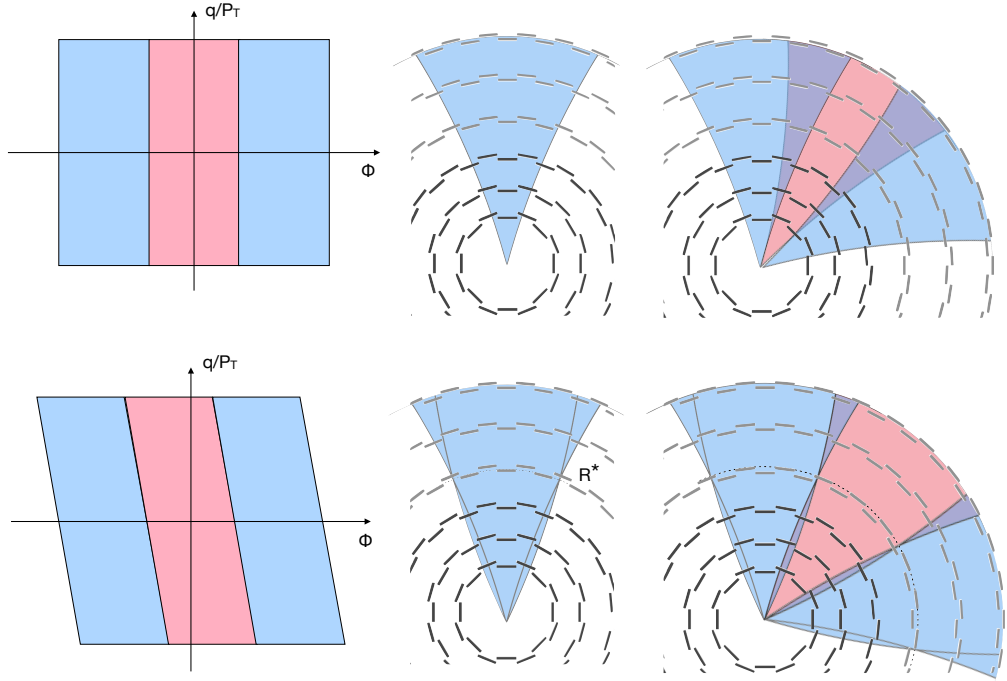}
\caption{Straight (top) and slant (bottom) definition of virtual trigger towers as seen in the transverse view. The cartoon on the left shows three adjacent towers in the $(\phi, q/\pt)$ parameter space. The cartoon in the middle shows how a single tower projects to the region of the detector traversed by tracks from the corresponding region of parameter space.
The slant definition is obtained by selecting the tracks on the basis of the $\phi$ at an optimized value of the radius $R^*$ instead of the $\phi$ at the origin. The sketch on the right shows how adjacent towers overlap in detector space. Stubs coming from purple regions need to be duplicated and sent to both trigger towers. The slant solution reduces the overlap between physical towers by a significant fraction with respect to the straight solution. $R^*$ is optimized to minimize overlap.}
\label{TriggerTowers}
\end{figure}

We decompose the four-dimensional track parameter space into the Cartesian product of two orthogonal two-dimensional spaces. The transverse space is $\phi$ versus $q$/\pt (which maps all high \pt values, for positive and negative charges, including infinite \pt, into a single connected segment) and the longitudinal space is $\eta$ versus $z_{0}$. The most straightforward way to divide the transverse parameter space into  non-overlapping regions is shown in Fig.~\ref{TriggerTowers} (top). A similar subdivision can be implemented in the longitudinal projection for $\eta$ versus $z_{0}$. 

The bottom of Fig.~\ref{TriggerTowers} shows a slightly different way of defining virtual towers. The difference is that the virtual tower a track belongs to is not determined by the $\phi$ of the particle at the origin, but by the $\phi$ of the particle at an optimized value of the radius~$R^*$. For our particular application, $R^*$ is optimized to achieve the minimum amount of overlap between adjacent physical towers, that is the minimum number of stubs that need to be duplicated and sent to multiple towers. Given the particular shape taken, in this case, by virtual towers, this solution is referred to as {\em slant} and reduces the overlap between physical towers by a significant fraction with respect to the straight solution. The cartoons in Fig.~\ref{TriggerTowers} show how the straight and the slant virtual towers translate to physical space and how the overlap regions are changed for the transverse projection. The shape of a single physical trigger tower is shown in the center cartoon, while the right cartoon shows how adjacent towers overlap. A similar improvement occurs for the longitudinal projection. 

The potential problem of track duplication caused by the same track being reconstructed in two different towers is virtually eliminated by the non-overlapping parameter space assigned to each tower. Each physical tower will in fact be able to discard any track reconstructed outside its own assigned space.

\subsection{Full Mesh Architecture}
Since the efficient data dispatching, for time and regional multiplexing, requires high bandwidth, low latency, and flexible real-time communication, a full-mesh architecture is a natural solution. In such an architecture, the detector data from any given trigger tower are received by a tower processor consisting of a set of pattern recognition boards (PRBs), all connected to each other via dedicated high speed serial links. The exact choice of the regional and time multiplexing factors is determined by considering the available bandwidth, processing power, and cost.

\begin{figure}
\centering
\includegraphics[width=12cm]{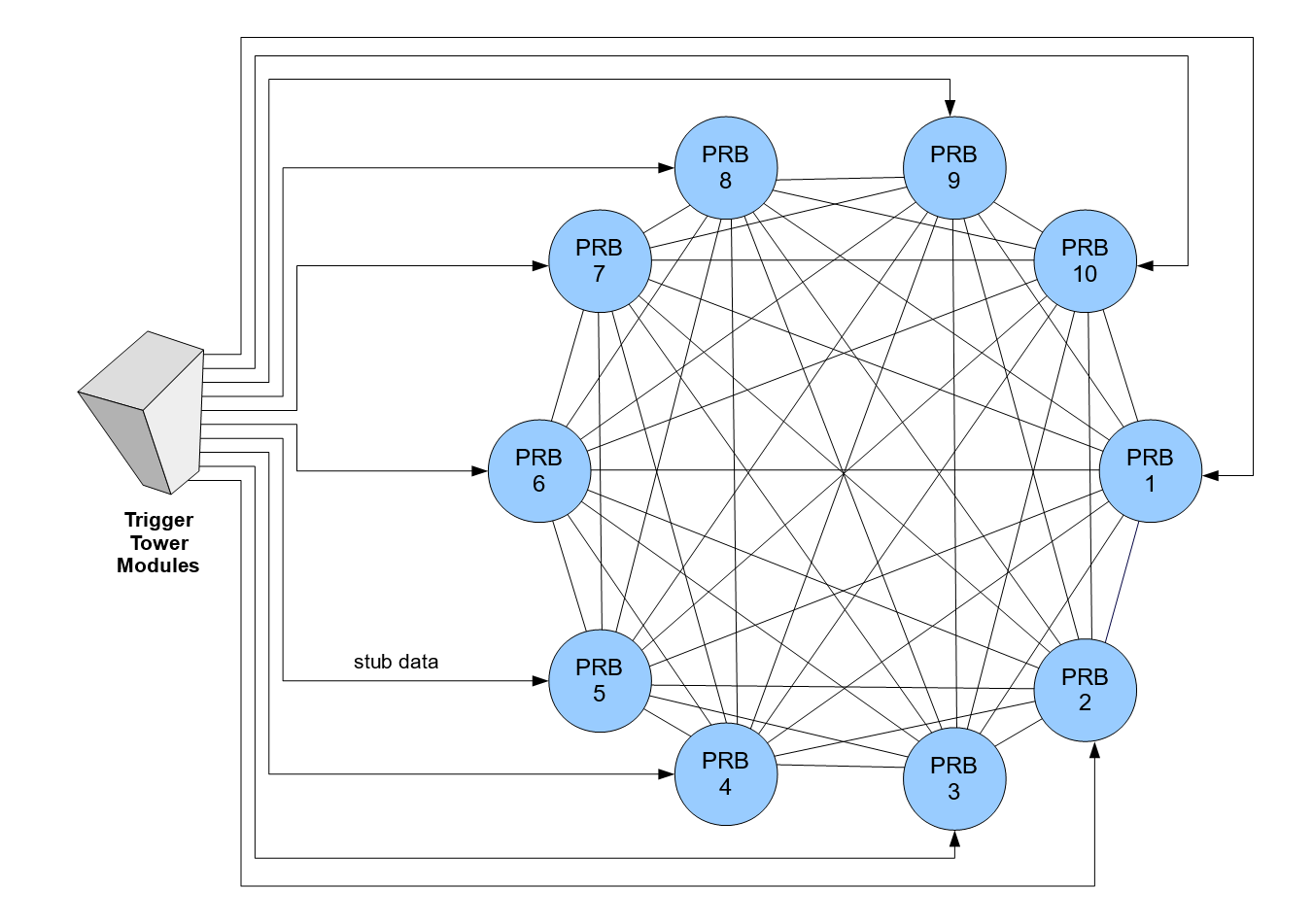}
\caption{Illustration of the data distribution to different Pattern Recognition Boards in the time-multiplexed round-robin scheme.}
\label{fig:dataflow}
\end{figure}

Such a full-mesh communication architecture can be implemented, in practice, either with optical fibers, or with a copper backplane. Each PRB supports one or more pattern recognition mezzanine (PRM) cards, connected to it via high-speed serial links, that function as pattern recognition and track fitting engines. Once the data has been unpacked by each PRB, they are then sorted by bunch crossing and transmitted over the full-mesh network to the proper PRB for additional processing by the target PRM. By coordinating the data transfers the PRBs work together to process event data in a time-multiplexed round robin scheme, as illustrated in Fig.~\ref{fig:dataflow}.

The system architecture described above satisfies the requirements of flexibility and scalability. It can accommodate a wide range of different regional and time multiplexing scenarios. For example, if the bandwidth capability of input and board-to-board communication links and the PRM processing speed were to double in the future, one could simply scale the size of the system down by a factor of two. Similarly, one could switch to a different multiplexing scheme and use two PRMs in each PRB to process data from the same event in parallel. It should be noted that the presented architecture is not necessarily tied to the AM component of this approach and can serve as a data delivery solution for a wide range of Level-1 reconstruction problems. The intrinsic benefit of the full-mesh concept is that it effectively minimizes the separation between processors, enabling system architects to experiment with different shelf configurations and to balance the input and board-to-board communication load between different processing units. 

The stubs delivered to the PRMs use local coordinates, measured relative to each detector module (e.g. module id, strip number, pixel number, etc.). They must be transformed to global coordinates, relative to the global common reference system (e.g. $\eta, \phi, r$), and stored in the Data Organizer. During this coordinate transformation, geometry and alignment corrections can be applied to the stubs using lookup tables.

\begin{figure}
\centering
\includegraphics[width=12cm]{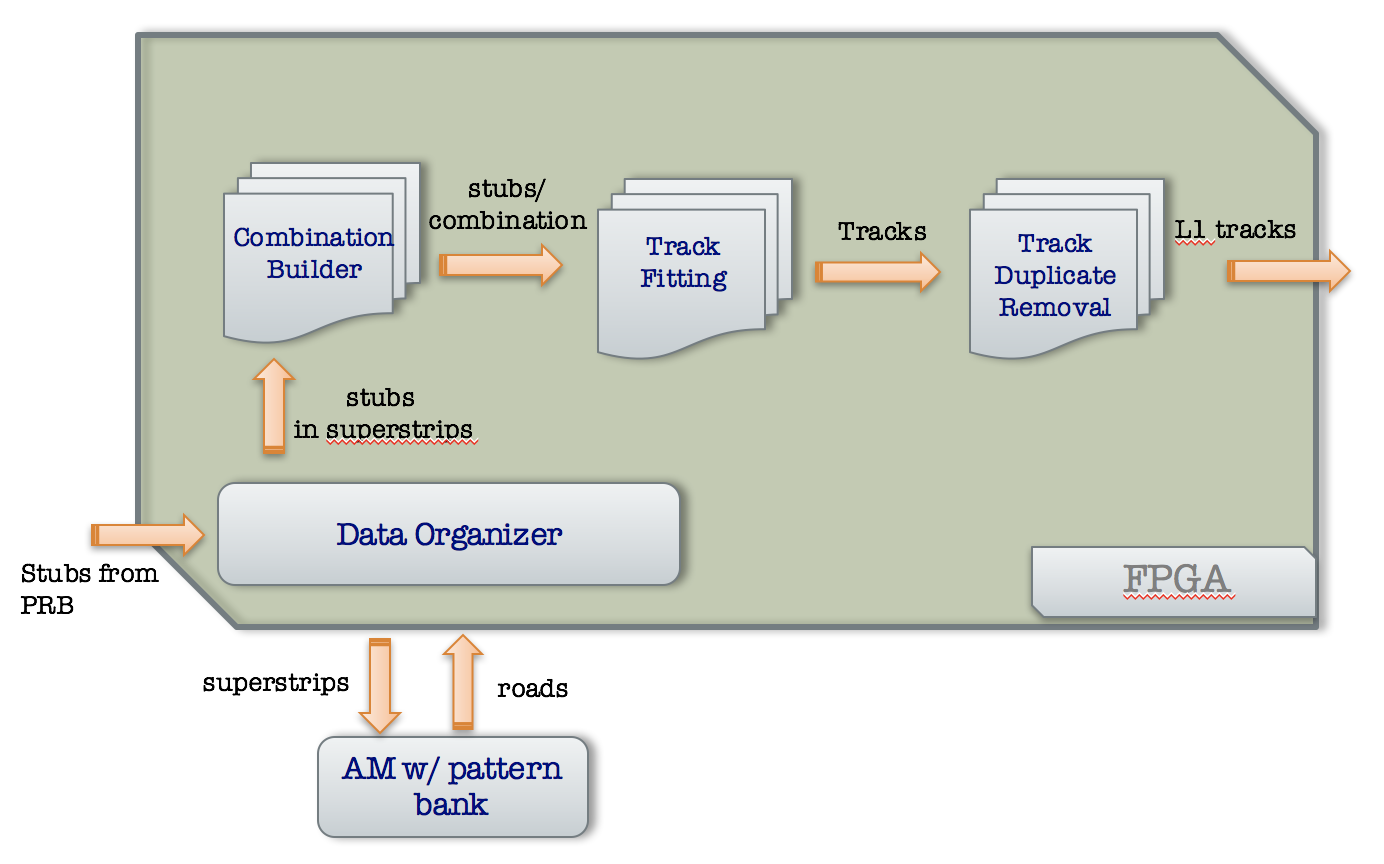}
\caption{The data flow and processing steps inside the PRM.}
\label{fig:prmflow}
\end{figure}

%% file: PatternRecognition.tex
\section{Pattern Recognition and Track Fitting\label{sec:PROptimization}}

Figure \ref{fig:prmflow} illustrates the data flow and processing steps inside the PRM. The processing steps include pattern recognition, track fitting, and track duplicate removal. A detailed description of each of these stages is provided in this section.

The pattern recognition stage is based on the use of Associative Memory. The AM concept was developed at the beginning of the 1990's for the CDF experiment~\cite{DellOrso:1988gmu}.
It was first implemented as an ASIC while, more recently, implementations in FPGAs have become practical, although with reduced pattern density if compared to an ASIC. For the demonstration purposes, we implemented AM functionality with reduced pattern density in an FPGA. This was sufficient in order to measure the system latency. Implementation of the full-size system relied on the assumption that an AM device with required pattern density becomes available either through a dedicated ASIC or a larger FPGA. 

\subsection{Superstrip Definition}
The operating principle of the AM requires the tracker silicon modules on all layers to be subdivided into a finite number of superstrips. Each superstrip contains a predetermined number of strips. Any combination of superstrips from different detector layers, compatible with the trajectory of a single charged particle above a certain \pt threshold, defines a valid pattern. The ensemble of all valid patterns constitutes the pattern bank that is loaded into the AM. 

In the particular problem we are addressing here, one of the most important design choices when searching for an effective solution using the AM approach is the choice of the size of the superstrips in the different layers of the tracker.

In a system with several layers, such as the CMS Phase-2 tracker, the number of patterns needed to capture all possible tracks grows as a high power of the number of superstrips, and thus with the inverse of the superstrip size, quickly exceeding the capacity of the AM. That sets a lower bound on the superstrip size. On the other hand, if superstrips are too large, the probability of having multiple stubs in the same superstrip will increase, requiring additional calculations to identify combinations of stubs belonging to the same valid track and discard accidental coincidences. 

The optimal superstrip size depends crucially on the occupancy of the detector, with higher occupancies requiring smaller superstrips and, consequently, a larger AM. It is important, however, for the superstrips not to be smaller than few times the detector's position measurement uncertainty. This would be of very little value as it would result in a large increase in the number of patterns with no significant improvement in the information carried by each pattern. 

From this discussion, one can see how the determination of the optimal superstrip sizes is a complex multidimensional optimization problem involving the type of particle that produced the track, the geometry, precision and occupancy of the detector, the pattern capacity of the AM, and the computing power available to solve the residual ambiguities from the stubs in the roads. Typical figures of merit in the optimization of a track trigger application are the total time needed to complete the track finding process, the  reconstruction efficiency, the fraction of fake tracks (purity), and the precision of the track parameters.

We start our optimization process with the determination of the minimum superstrip size compatible with the position measurement errors. In our case, the main contribution to the measurement error is from multiple scattering caused by the increasing amount of material that charged particles traverse as they travel from  inner to  outer detector layers. A detailed simulation of the detector is used to generate a large number of muon tracks and compare the position of the stubs recorded in each layer to the position where the hit would ideally be if there were no measurement errors or multiple scattering. Tracks are generated with uniform distributions in pseudorapidity and $1/\pt$, and with a Gaussian distribution of the primary vertex position along the beam direction with $\sigma = 5$\,cm, as expected for the HL-LHC running conditions. The minimum track \pt value is set to $3$\,GeV.

\begin{figure}
\centering
\includegraphics[width=15cm]{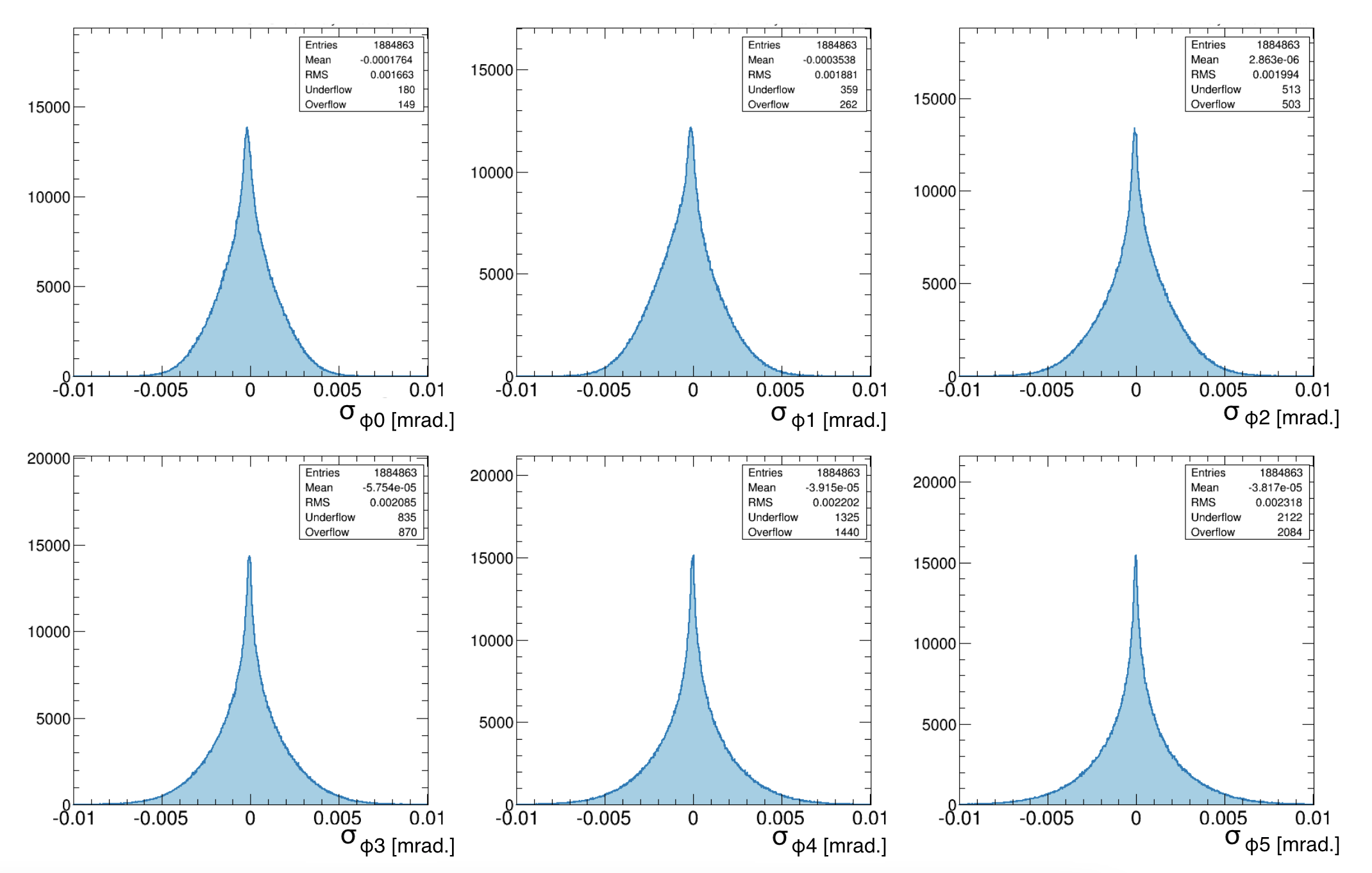}
\caption{The distribution of the difference between the actual and the ideal hit position along the $\phi$ direction used in the definition of the superstrip width. $\phi_0$ corresponds to the innermost layer and $\phi_5$ to the outermost one.}
\label{fig:sswidth}
\end{figure}

The distribution of the difference between the actual and the ideal position along the $\phi$ coordinate ($\sigma _{\phi}$) is shown in Fig.~\ref{fig:sswidth}, for each detector layer separately. Each distribution was integrated symmetrically, starting from the center, to include 90\% of its area and the resulting interval (typically of the order of few mm) is taken as a baseline width for the superstrips along the $\phi$ coordinate for the corresponding layer. As we go from the inside of the detector to the outside, due to the increasing amount of material traversed, the size of the superstrips is increasing in terms of the angle as seen from the origin. It is because of this feature that this particular superstrip configuration is referred to as {\em fountain}. 

We generated different fountain configurations by applying a scale factor (sf) to multiply the width of all the superstrips while maintaining a constant ratio between different layers. We also tested a further subdivision of the detector layers along the $z$ direction, resulting in two-dimensional superstrips. The number of divisions in $z$ is denoted as nz. The results of this sub-division optimization are discussed below.

\subsection{Generation of the Pattern Bank}
To generate a pattern bank, we also need to define the target parameter region ($\eta$, $\phi$, \pt and primary vertex position in $z$) for which we want to maximize the track finding efficiency. Once the superstrips and the target parameter region are specified, the procedure to generate the pattern bank is as follows. For each track, in a sufficiently large simulated sample from the target parameter region, one records which superstrip it goes through in each detector layer. A pattern is defined as the set of such superstrips, one per detector layer, each with a unique integer identifier. The ensemble of all such different patterns generated by all the tracks in the sample constitutes the pattern bank.

\begin{figure}
\centering
\includegraphics[width=10cm]{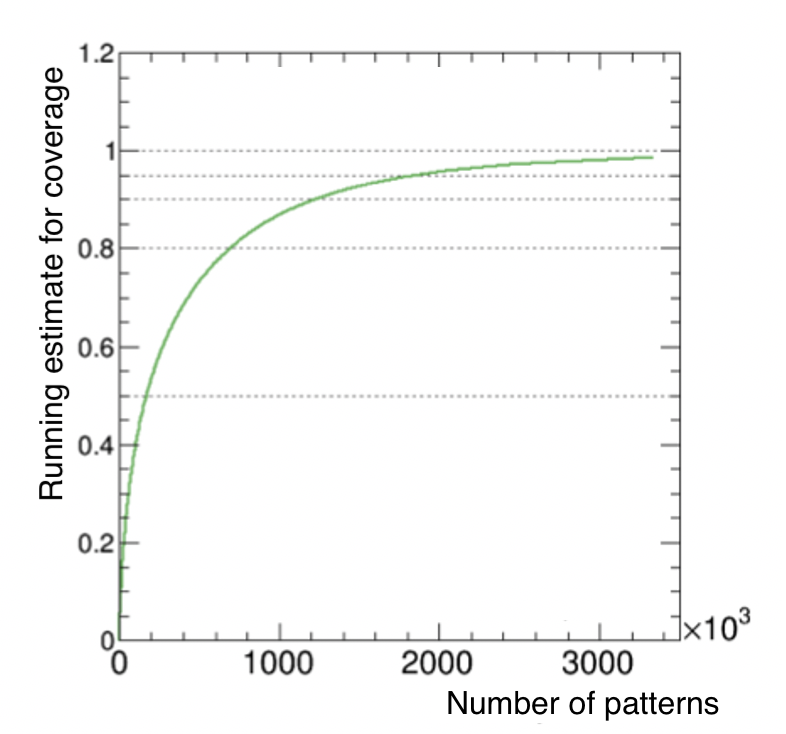}
\caption{Coverage as a function of the number of patterns included in the bank.}
\label{fig:coverageVspatterns}
\end{figure}

The coverage of a pattern bank for a given superstrip configuration is defined as the probability that a track, randomly extracted from the target parameter region, generates a pattern that is included in the pattern bank. In practice, it is not possible to reach 100\% coverage; however, one can get closer and closer as one generates more and more patterns. Figure~\ref{fig:coverageVspatterns} shows a typical plot of the pattern bank coverage as a function of the size of the pattern bank. The plot shows how the curve tends to converge to full coverage as more and more patterns that are less and less probable are added to the bank.

During the pattern bank generation process, the same patterns are being found multiple times. The number of times a pattern has been found provides an estimate of the probability for a particular pattern to be produced by a track randomly extracted from the target parameter region. We refer to this probability as the pattern's popularity. The strategy we pursue to optimize the coverage of the pattern bank, given the limitation of the AM size, is to first generate a very large pattern bank and then include, in decreasing order of popularity, only the most popular patterns until we filled all the available AM. The pattern bank can also be optimized differently. For example, if tracks with larger \pt are of higher importance (which is often the case), one could sort patterns by the average \pt of the tracks that generate that pattern.

An important feature of the pattern bank is that it can be stored in the AM in any order. The particular order chosen at time of loading the AM will also be the order in which the roads appear at the output at the end of the pattern recognition phase. This is useful if, for example, latency considerations require truncation of the number of roads that can be processed. For instance, if patterns are sorted by decreasing \pt, and stored in the AM in that particular order, the roads will go from the AM into the track fitting stage starting from the highest \pt and progressively to lower \pt. This can be useful in preserving the trigger efficiency for high \pt tracks in particularly complex or busy events.


\subsection{Pattern Bank Optimization}
For any given assumption on the properties of the collisions in the detector, and for a given desired coverage, the different superstrip configurations will yield a different number of patterns needed in the pattern bank. Correspondingly, a different number of roads will be fired on average during data taking.
Then, for each road, a different number of stubs will appear in each  layer. To find tracks, the track fitting algorithm needs 
to combine stubs from different layers and the more stubs in each layer, the more combinations it needs to process. Then some criteria, such as the best goodness of fit, will need to be applied in order to pick the best track among all combinations. Therefore, in addition to the pattern bank size and the number of roads, a figure of merit that addresses the resource usage and latency of the track fitting process in the FPGA is the number of combinations to fit, given by the product of the number of stubs in each layer of the roads fired, summed over all roads.


The studies described here focus on a single trigger tower with a pseudorapidity coverage of $0 < \eta < 1.0$. The detector is symmetric in $\phi$, so the  position in $\phi$ of the tower is not important. Moreover, simulation studies not addressed in this paper indicated that the conclusions and trends obtained in the central tower studies also apply to the towers at higher $\eta$, with the caveat that a larger number of patterns (up to 50\% more) per trigger tower may be needed to achieve the desired coverage. For the studies presented here, we use PU200 samples. 

\begin{figure}
\centering
\includegraphics[width=10cm]{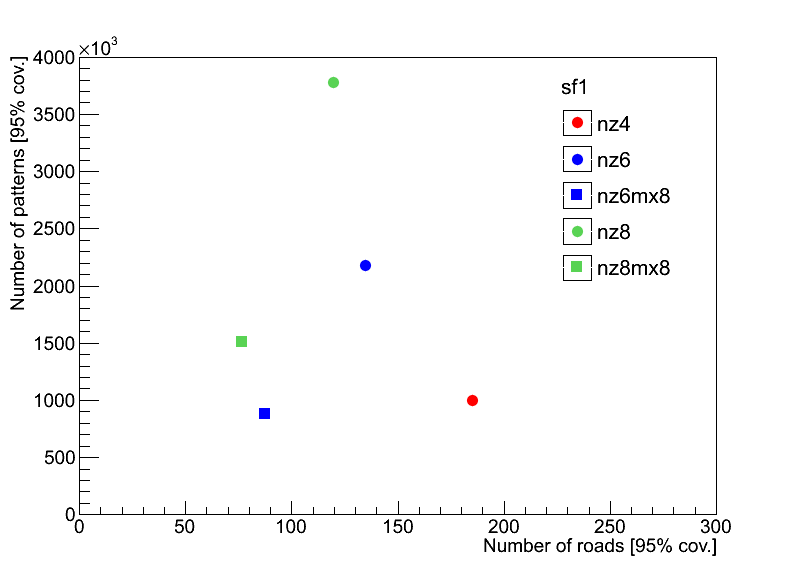}
\caption{The number of patterns (with 95\% coverage) versus the average number of roads per event for different superstrip configurations used in the optimization of the AM+FPGA approach. Configurations closer to the origin are more optimal as they correspond to a smaller bank size and lower road multiplicity.}
\label{fig:patternsVSroads}
\end{figure}
\begin{figure}
\centering
\includegraphics[width=10cm]{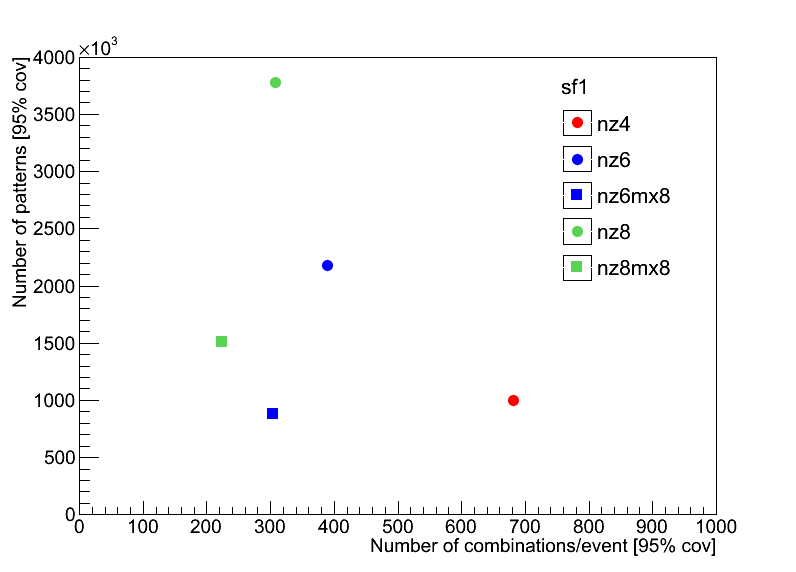}
\caption{The number of patterns (with 95\% coverage) versus the average number of combinations per event for different superstrip configurations used in the optimization of the AM+FPGA approach. Configurations closer to the origin are more optimal as they correspond to a smaller bank size and lower number of combinations.}
\label{fig:patternsVScombs}
\end{figure}

Some patterns including adjacent superstrips can be merged into a single one if Gray code is used for the sequential superstrip number in conjunction with three-state bits~\cite{Annovi:2228284}. Figure~\ref{fig:patternsVSroads} shows, for different superstrip configurations, the total number of patterns in the bank (for 95\% coverage) as a function of the average number of roads generated in each crossing. The sf1 refers to a scale factor of 1.0 in the fountain configuration, the number of divisions along the $z$ direction varies from nz4 to nz8, with nz4 corresponding to the larger superstrip sizes, and mx8 refers to a configuration in which merging of up to eight patterns with adjacent superstrips is performed. Figure~\ref{fig:patternsVScombs} shows the number of patterns as a function of the total number of combinations as described above. One can clearly see the interplay between the size of the pattern bank and the number of roads and combinations. The better configurations are those in which both the number of patterns and the number of roads and combinations are smaller, as this will be implementable with a reasonably sized AM and a reasonably powerful FPGA. One can also clearly see that there are significant differences between the configurations, showing how important this optimization process really is and how much gain can be obtained from it. 
 
Based on these studies, sf1nz8mx8 is chosen as the baseline pattern bank configuration. It provides the smallest number of roads and combinations as well as a bank size targeted by the ongoing AM R\&D efforts. This configuration requires a pattern bank consisting of about 1.5 million patterns which will result in up to 70 roads and 200 combinations per trigger tower per event in the PU200 conditions. At the time of the work presented here, the largest AM chip available was the AMChip06~\cite{Annovi:2228284}, which featured 128,000 patterns and was designed for the ATLAS FTK, a Level-2 trigger system~\cite{Shochet:1552953}. Demonstration results using that chip can be found in Ref.~\cite{Fedi:2018ewd}. Two parallel efforts were ongoing with the goal to develop an AM chip with the pattern density and speed adequate for L1. The Vertically Integrated Pattern Recognition AM (VIPRAM) project~\cite{Hoff:2017vib} relied on the 3D Very Large Scale Integration (VLSI), while the AMChip~\cite{Annovi:2017suq} utilized planar VLSI 28\,nm technology. Both developments aimed to achieve over 150,000 patterns per chip and the required operational speed. Eight to ten of such ASICs per PRM would be sufficient to accommodate the baseline bank. Reconstructing tracks with $\pt>2$\,GeV (as opposed to the nominal $\pt>3$\,GeV) would result in an approximately 30\% increase of the pattern bank size. 


\subsection{Track Fitting \label{sec:TrackFitting}}
Following the pattern recognition stage in the AM, the stubs in the matched roads are used as input to the track fitting stage. The fitting procedure is modeled very closely to the one originally used for the SVT trigger in the CDF experiment~\cite{Ashmanskas:2003gf}.

The first task of the track fitting stage is to discard the largest possible fraction of random stub combinations by using their full resolution position coordinates and thus improving the precision of the pattern recognition. In the approximation of infinite detector resolution, the requirement that all the stubs in a set belong to a single track can be expressed as $F_i(x_k) = 0$, where $x_k$ is the coordinate of the stub in the $k^{th}$  detector layer and $F_i$ is a set of geometrical constraints. The effect of finite detector resolution is to make the $F_i$'s random variables taking values only slightly different from zero. To solve the problem of track finding, for each candidate track, we evaluate the $F_i(x_k)$ and accept the track if and only if the $F_i$'s are within the appropriate boundaries. In general, $F_i(x_k)$ are difficult to compute exactly, so we substitute them with their linear expansions around some convenient points $x_{0k}$, one for each road and each detector layer:

\begin{equation}
F_i(x_k)\simeq {\left(\partial F_i\over \partial x_k\right)}_{x_{0k}}  \cdot (x_k -x_{0k}) = v_{ik} \cdot x_k + c_{ik}.
\end{equation}

The stubs that fall within one road typically span a very limited coordinate range and linear approximation works very well. All the constants $v_{ik}$ and $c_{ik}$ can be determined from the detector geometry, either analytically or numerically. 

A numerical method, based on Principal Component Analysis (PCA)~\cite{Jolliffe2011}, which also allows for an easy handling of detector misalignment, is described in Ref.~\cite{Ristori:2010zz}. A similar method can be used to compute a first order approximation of the functional relationship between the position coordinates of the stubs and the values of the parameters of the track. The method involves simple linear algebra, performed by matrix multiplication, and allows to refine the pattern recognition, remove most of the random stub combinations, and compute track parameters.

Track fitting is performed with only one stub per tracking layer. However, due to the coarseness of the AM superstrips, especially in high occupancy events, multiple stubs per layer can exist within a given road. For each road, the Combination Builder forms all possible combinations of stubs, selecting at most one stub per layer. To preserve tracking efficiency in case one stub is not registered by the detector, combinations with stubs in five out of the possible six layers are allowed. The stub combinations produced by the CB are passed to a linearized track fitting processor for the track parameters determination. The matrix multiplication operations involved in track fitting can be executed with low latency using Digital Signal Processors (DSP) available in modern FPGAs. 

In a simple linearized track fit, the track parameter resolution is typically degraded due to non-linearities intrinsic in the detector geometry. The effect of non-linearities can be mitigated by reducing the size of the region of validity of each particular set of PCA coefficients. A different method is to increase the power of the expansion from first order to a higher order. Along this line, a novel approach has been developed to correct stub coordinates before the fit in order to partially correct for such non-linearities and to simulate a "flatter" detector. Reducing detector non-linearities results in a significant reduction in the number of constants needed for track fitting and results in improved tracking performance. This strategy has been proven to work very well and is described in detail in~Ref.\cite{Clement:2018apn}. 

\subsection{Duplicate Removal \label{sec:Duplication Removal}}

The main cause for the creation of track duplicates comes from stub combinatorics. As explained above, in order to obtain high tracking efficiency, we need to accept stub combinations from five detector layers out of the available six. This will recover tracks where one stub has been missed or mismeasured, but will also create five track duplicates for each good track with six good stubs. Also, occasionally, one superstrip will contain one noise hit in addition to the good one which will produce a duplicate track with sufficiently good fit to pass the cut. Another possible cause for track duplicates is if the same track is reconstructed in two different trigger towers. The probability for this to happen has been minimized by the construction of minimally overlapping trigger towers described in Sec.~\ref{sec:DataDelivery}. A given set of track parameters are contained only in one virtual trigger tower and, therefore, a good pattern for that track will be present only in that trigger tower. Unfortunately, the superstrips, being of finite size, create the possibility of a small overlap between patterns of adjacent towers. 

The fraction of duplicate tracks created by the system is large such that more than 80\% of tracks are duplicates. Our strategy is to have a dedicated stage, immediately after track fitting, to remove as many of them as possible. We have tested two ways of performing duplicate removal: one based on the number of stubs shared by the two tracks, the other based  on comparing the helix parameters of the two tracks within the corresponding resolutions. Each method has advantages and disadvantages. The second one relies on the track parameter resolutions, which vary depending on the particle type producing the track: resolutions for electrons are significantly worse than those for pions, which in turn are worse than for muons. Since one has no way to perform particle identification in the track trigger system, the choice of resolution parameters is non trivial. The first method is less ambiguous, however it requires carrying forward the information on the stub-to-track association through the track fitting firmware. This requires additional resources, but is not prohibitive with modern FPGAs. 

%% file: Demonstration.tex
\section{Demonstration System \label{sec:demo}}
In the following, we describe a demonstration system used to measure the latency and performance parameters of the AM+FPGA L1 tracking approach. 

\begin{figure}
\centering
\includegraphics[width=12cm]{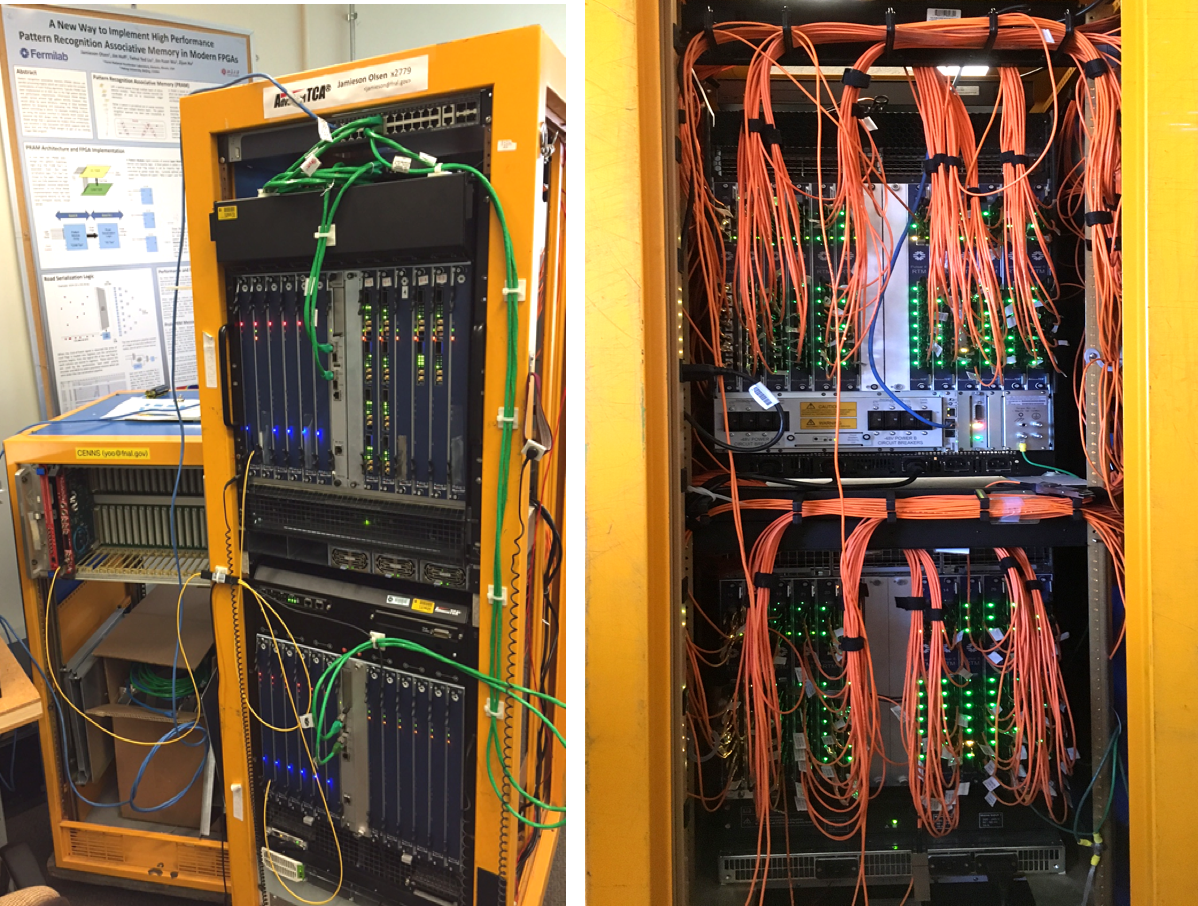}
\caption{Demonstrator hardware rack. The Data Source boards are in the lower ATCA shelf and the PRBs are in the upper shelf. The rear view of the rack (right) shows 120 QSFP+ optical cables connecting the two shelves.}
\label{fig:demosystem}
\end{figure}

A tower processor must support a large number of fiber transceivers and be big enough to host many pattern recognition engines. Given these requirements, a full-mesh ATCA shelf is chosen for the tower processor. For the demonstration system described in this paper, regional and time multiplexing factors of 48 and 20 are chosen, respectively. Figure~\ref{fig:demosystem} shows the primary hardware configuration used in the demonstration. The hardware consists of two ATCA shelves, each occupied by twelve Fermilab Pulsar IIb  ATCA boards. One ATCA shelf with 10 PRB boards is designated to process a trigger tower, with each PRB hosting two PRMs. There are therefore 20 PRMs each acting as an independent time multiplexed unit and receiving new data every 500\,ns. The second ATCA shelf includes boards used to emulate the transmission of front-end stub data from the tracker front-end electronics. These are referred to as Data Sources (DSs). 

\begin{figure}
\centering
\includegraphics[width=12cm]{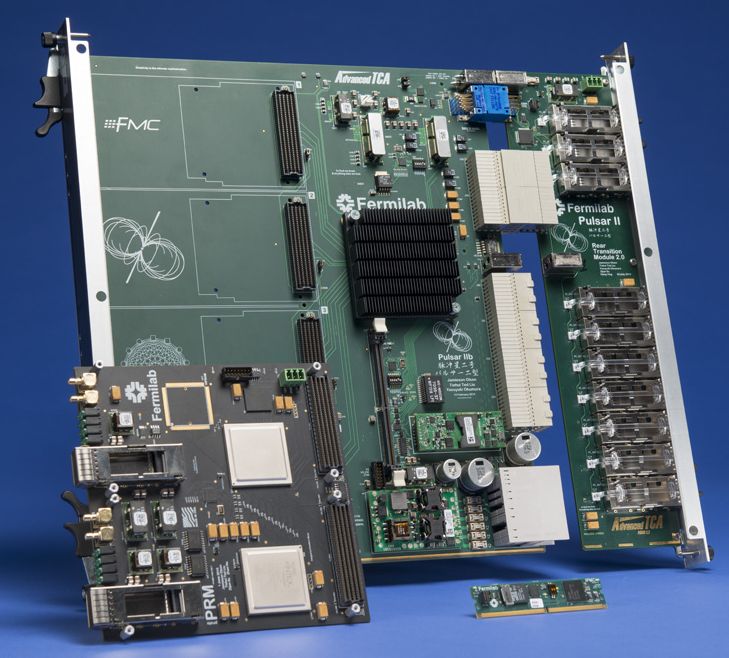}
\caption{The Pulsar IIb board and the PRM mezzanine card. The intelligent platform mezzanine card (IPMC) and the RTM described in Appendix A are also shown.}
\label{pulsar2}
\end{figure}

The Pulsar IIb and PRM hardware components are shown in Figure~\ref{pulsar2}. The Pulsar IIb board has been developed specifically for this project. As explained above, ten boards in one shelf were sufficient to emulate a trigger tower. The additional two boards are included in order to support mezzanine cards that are used for the clock and control distribution and for testing the performance of the ATCA backplane links between all slots in the shelf. A detailed description of the demonstrator hardware can be found in Appendix~\ref{sec:hardware}.

Ten Pulsar IIb boards are used as DSs in the demonstration. Each DS board sends its data over forty links to one of ten PRBs in the other ATCA shelf. The DS boards and PRB boards are synchronized to a CERN TTC optical link~\cite{Taylor:1998sx}. This TTC optical link originates from a TTCci VME board (on the left side of Fig.~\ref{fig:demosystem}) and is then passively split and sent to a pair of TTC receiver FMC mezzanine cards on the additional Pulsar IIb boards in the ATCA shelves. The link carries a 40\,MHz LHC clock and a special signal (B0), upon the arrival of which the DS boards start to send data out. The arrival time of the B0 signal into the DS shelf defines the starting point in the system latency measurements. The end point for the latency measurements corresponds to the time when the last track is reconstructed. 

\begin{figure}
\centering
\includegraphics[width=15cm]{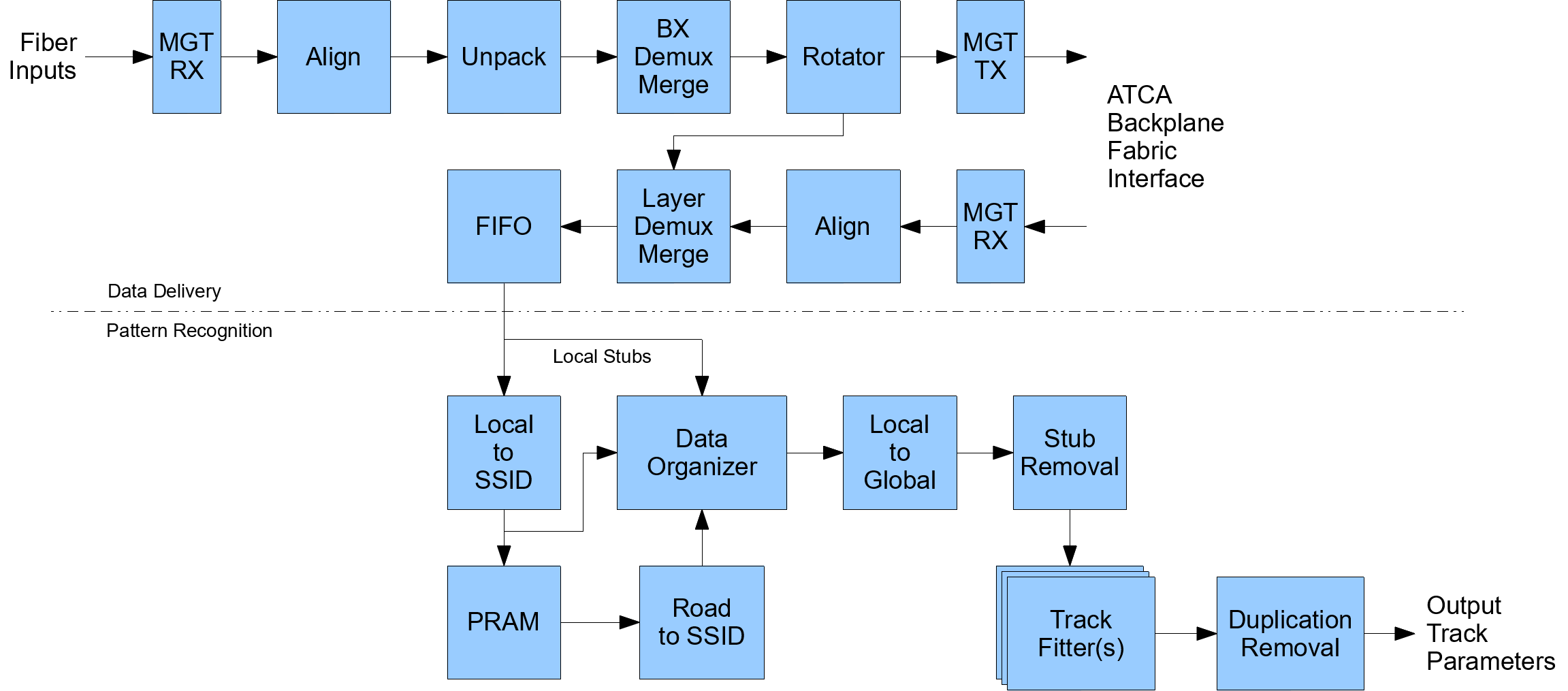}
\caption{Overall firmware structure used in the AM+FPGA approach. The top section corresponds to the data delivery performed in the PRBs, while the bottom portion describes the steps executed in the PRMs. MGT TX and RX indicate multi-gigabit transmitter and receiver, respectively. SSID stands for the superstrip ID.}
\label{prb:fw}
\end{figure}

The overall structure of the firmware is shown in Fig.~\ref{prb:fw}. Stubs arrive into the PRB over multiple high speed fiber optic links from upstream modules. Each PRB must first determine whether the stub data it receives belongs to one of the bunch crossings it is assigned to. If this is so, the stubs must be formatted and sent to the designated PRM for processing. If not, the stubs must be formatted and routed to the appropriate PRB. Stub communication between the PRBs proceeds via the backplane. Each PRB accepts all the stubs it receives both through its Rear Transition Module (RTM) and through its backplane, and formats and delivers them to their target PRMs. 

For the demonstration, a fully pipelined and cycle-accurate emulation of AM operations runs in one of the PRM \mbox{FPGAs}, while the other implements the DO, CB, and TF functionalities. In the demonstrator firmware, 1024 patterns were used in the FPGA emulation. This suffices for the demonstration of AM+FPGA latency and efficiency, because even in a full-scale system, only a small number of patterns will be matched in a given bunch crossing.

All firmware corresponding to the DO, CB, and TF steps is fully pipelined. The design of the DO component utilizes newly supported features in Xilinx block RAMs to provide low latency stub storage and retrieval. The track fitting pipelines are implemented using cascading DSPs and distributed RAM resources for the matrix coefficients in order to minimize latency and resource usage. In the baseline scenario, the DO distributes roads to four TF units which run in parallel.

%% file: Results.tex
\section{Results \label{sec:results}}
\subsection{System Latency}
The latency is measured using the two-shelf demonstrator system described in the previous section. The system includes the firmware implementation of nearly all of the FPGA tracking logic, only the duplicate removal step is not implemented.
However, estimates based on the number of required operations suggest that the resource usage and latency introduced by this step is very small compared to the other stages of the data processing chain. Intermediate latency measurements are performed for some of the most relevant internal processing stages. 

A summary of the latency of the system, broken down into the main processing stages, is shown in Table~\ref{tbl:latency}. 
Measurements were performed using an oscilloscope connected to appropriate internal firmware counters. Some of the shorter processing stages are not included in the table. Due to the pipelined nature of the firmware, some of the processing stages overlap in time. All data transmission latencies are included. Data sourcing and delivery latencies include all operations needed for L1 tracking, including those needed for data sorting.

\begin{table}
\centering
\begin{tabular}{|c |  c | c| }
\hline
Processing stage  &  Begin (ns) & End (ns) \\ \hline
Module data arrives at RTM    &  0  &  200 \\
Stub sorting by bunch crossing&  330 & 530 \\
Stub transfer over backplane  &  630 & 830 \\
Stub sorting by detector layer&  960 & 1460\\
Stub transfer from PRB to PRM &  1085& 1585\\
Convert coordinates to superstrips   &  1215& 1732\\
Output of AM roads           &  1773& 2273\\
Output of DO stubs           &  1840& 2340\\    
Stubs ready for CB+TF         &  1881& 2381\\
Tracks out of TF array        &  2056& 2556\\
\hline
\end{tabular}
\caption{The latency of the system broken down into the processing steps. Begin and end times for each stage are given. Processing stages with less than 200\,ns are not shown in the table. Due to the pipelined nature of the firmware, some of the processing stages overlap in time. The total latency is approximately 2.5\,$\mu$s.}
\label{tbl:latency}
\end{table}

The overall latency corresponding to the design with four track fitting engines running in parallel is approximately 2.5\,$\mu$s. The tracking performance corresponding to this latency is presented later in this section. Configurations with twice and thrice the number of track fitting engines increase the total latency to approximately 3.0 and 3.5\,$\mu$s, respectively, but allow to maintain excellent track reconstruction performance inside high energy jets, where the local stub occupancy is very high. The  configuration with 4 TFs is implemented in the Xilinx KU060 FPGA with comfortable resource usage. Configuration with 8 TFs occupies most logical cells in the KU060 and the one with 12 TFs requires more cells than available in the FPGA. However, this and even more demanding schemes should be implementable with more modern FPGA devices. 

\subsection{Tracking Performance}
The performance of the system is studied using large sets of simulated events in the framework with floating point precision. The framework includes simulation of the duplicate removal step. Bit-wise hardware emulation is used to assess the impact of the limited fixed point precision used in the firmware as well as any truncation effects originating from the limited bandwidth of the system. Bit-wise emulation is required to perfectly match the demonstrator system output. Perfect agreement is observed using a representative sample of approximately 10,000 events. Truncation effects are introduced in the simulation via equivalent cuts on the stub, road, combination, and track multiplicities. 

The following  parameters are considered in the evaluation of the system performance: (i) track finding efficiency as a function of transverse momentum and pseudorapidity for different types of charged particles (muons, electrons, pions); (ii) resolution of reconstructed track parameters (\pt, $\phi_{0}$, $z_{0}$, $\eta$) as a function of transverse momentum and pseudorapidity for different types of charged particles; and (iii) the purity of tracks reconstructed by the system. It should be noted that the absolute value of efficiency and purity depends on the definition of real, fake, and duplicate tracks. In the quoted numbers, very tight matching requirements were used to associate reconstructed tracks to simulated particles. Namely, in order for a track to be called "true", all stubs used in the reconstruction need to originate from the corresponding simulated particle. 

\begin{figure}
\centering
\includegraphics[width=12cm]{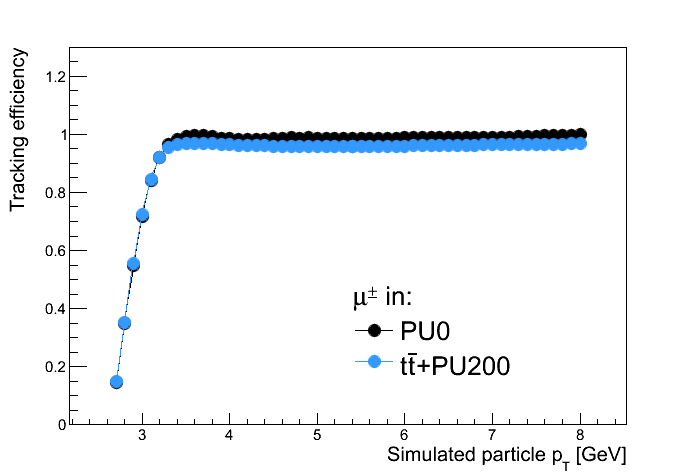}
\includegraphics[width=12cm]{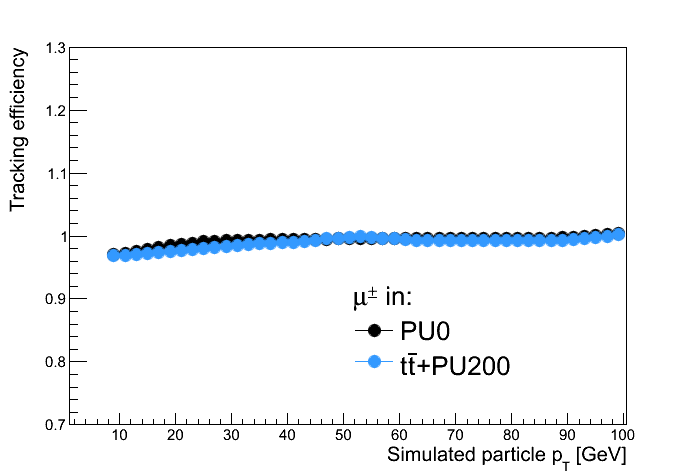}
\caption{Efficiency of L1 tracking for muons with low (top) and high (bottom) transverse momentum. For comparison the curves are shown for isolated muons without pileup and for muons in $\mathrm{t\bar{t}}$+PU200 events. Statistical uncertainties are small.}
\label{fig:eff_pt}
\end{figure}

\begin{figure}
\centering
\includegraphics[width=12cm]{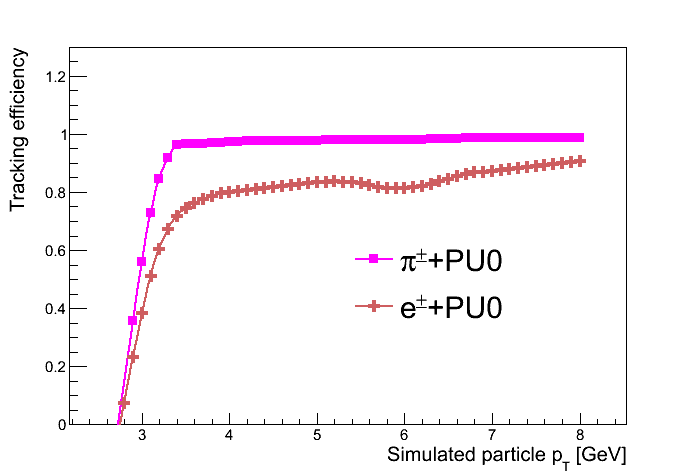}
\caption{Efficiency of L1 tracking for pions and electrons with low transverse momenta. The curves are evaluated in the absence of pileup. Statistical uncertainties are small.}
\label{fig:eff_pt_epi}
\end{figure}

\begin{figure}
\centering
\includegraphics[width=12cm]{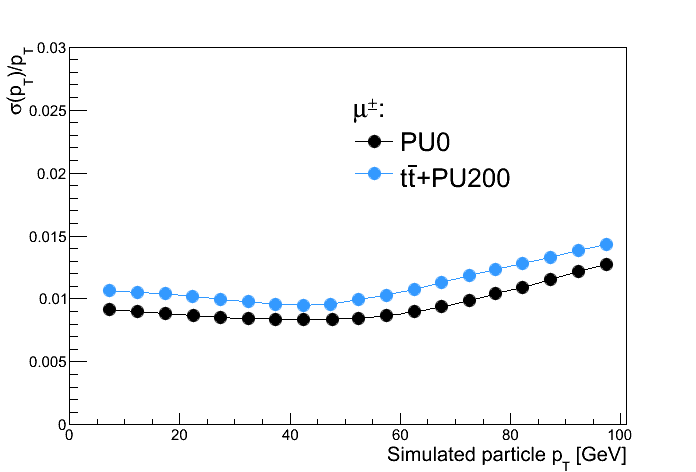}
\includegraphics[width=12cm]{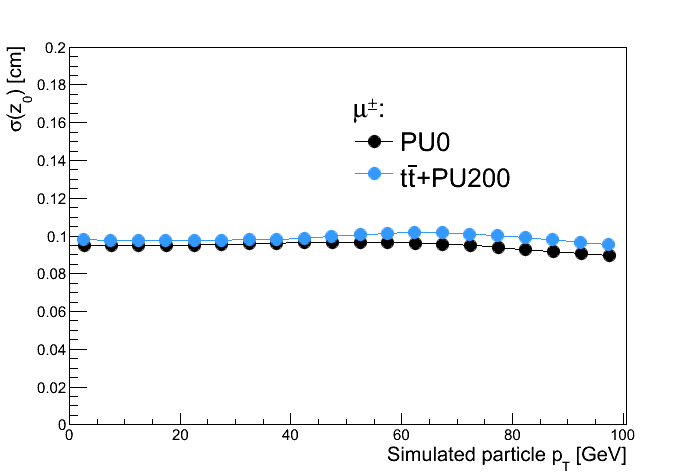}
\caption{ Relative \pt and absolute $z_{0}$ resolutions for single muons as a function of their transverse momentum. For comparison the curves are shown for isolated muons without pileup and for muons in $\mathrm{t\bar{t}}$+PU200 events. Statistical uncertainties are small.}
\label{fig:res_pt}
\end{figure}

Evaluating efficiency and resolutions of a single track reconstruction in the absence of pileup is essential for quantifying the performance of the system and for understanding its fundamental limitations. The tracking efficiency as a function of \pt for muon tracks is shown in Fig.~\ref{fig:eff_pt}. The muon reconstruction efficiency curve shows excellent performance, has a sharp turn-on near 3 GeV, and flattens at approximately $99\%$. The efficiencies of reconstructing the input tracks as a function of \pt for pion and electron tracks are shown in Fig.~\ref{fig:eff_pt_epi}. The efficiency for pions is slightly worse than for muons due to larger effects of multiple scattering in the detector, but is still above $95\%$. For both muon and pion tracks the efficiency remains flat in the studied range of up to 100\,GeV. The electron efficiency has a much slower turn-on due to bremsstrahlung effects and reaches a plateau of about 90\% at 10\,GeV. The effect of photon radiation alters the trajectories of the electron tracks and make them fail both the pattern recognition stage and the track fit quality requirements. Potential improvements in electron reconstruction can be expected from generating dedicated patterns for electrons and retaining tracks with higher normalized $\chi^{2}$ values, however no such attempts have been made in the scope of this work. For the muon and pion tracks, the typical relative \pt resolution is at the 1\% level, $\phi$ and $\eta$ resolutions are at the level of $10^{-3}$, and the $z_0$ resolution is approximately 1\,mm. The resolutions remain mostly flat as a function of the track transverse momentum and pseudorapidity up to $|\eta|<1.0$. Example resolutions are shown in Fig.~\ref{fig:res_pt}. More details, including the track fitting performance in the forward region of the detector, can be found in Ref.~\cite{Clement:2018apn}.

\begin{figure}
\centering
\includegraphics[width=12cm]{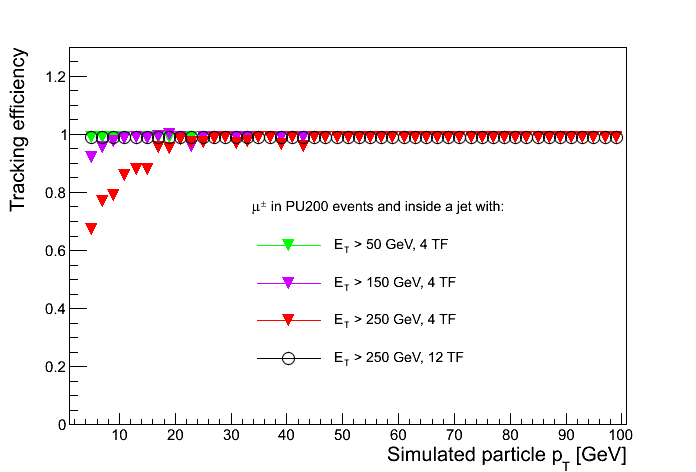}
\caption{Efficiency of L1 tracking for muons inside jets of different minimum transverse energy \Et (50, 150, and 250\,GeV) in events with PU200. Results were obtained with the default demonstrator configuration with 4 TFs. The performance of the configuration with 12 TFs is also shown for jets with \Et$>$\,250 GeV.}
\label{fig:eff_jets}
\end{figure}

Pileup will adversely impact both the pattern recognition and the track fitting stages of the track reconstruction. There are several ways in which stubs originating from pileup interactions can
impact the efficiency of finding tracks. Firstly, these stubs can substantially increase the number of roads and combinations in the pattern recognition stage of the data processing. In this situation, and due to limited time available for each stage of the processing chain, some true roads and combinations may not be processed, leading to inefficiency. Secondly, combinatorial stubs originating from PU interactions may contaminate real combinations, leading to high  $\chi^{2}$ values of the track fit, again resulting in an inefficiency. Studies were performed to determine the efficiency of finding isolated tracks in $\mathrm{t\bar{t}}$ events and under various pileup conditions. It is found that the performance of the system in terms of track finding efficiency and resolution remains essentially intact in \ttbar+PU200 events, as shown for muons in Fig.~\ref{fig:eff_pt} and Fig.~\ref{fig:res_pt}, respectively. The effect of pileup on pion and electron reconstruction is found to be similar in size. Another question to ask is how well we can reconstruct tracks inside jets of transverse energy larger than 50\,GeV. The conditions inside the jets are characterized by very high local stub density, leading to additional combinatorial challenges. In the absence of truncation effects, the efficiency is found to be the same as for isolated muons and pions. However, when truncation effects are taken into account, an inefficiency appears for low momentum tracks and increases as a function of jet energy, as shown in Fig.~\ref{fig:eff_jets}. This inefficiency originates primarily from the limited number of roads that can be processed within the latency requirements. Figure~\ref{fig:eff_jets} demonstrates that this effect can be mitigated by the additional processing power available in configurations equipped with 12 TFs.

\begin{figure}
\centering
\includegraphics[width=12cm]{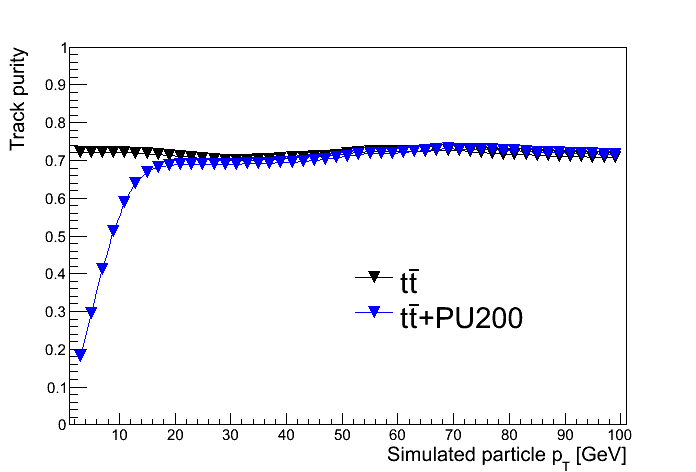}
\caption{Purity of the L1 tracks as a function of their transverse momentum in \ttbar events. The purity is largely constant at approximately 70\% with the exception of the low \pt part of the distribution, where the purity reduces to 20\% at PU200.}
\label{fig:purity}
\end{figure}

We require that the system produces high purity tracks while maintaining high reconstruction efficiency. This is desirable as a large fraction of fake or duplicate tracks in the output sample would significantly complicate the downstream processing of the data. The average number of tracks above 3~GeV in the entire detector is found to be approximately 55 (75) per collision for \ttbar+PU140 (PU200). The purity of tracks in \ttbar events (shown in Fig.~\ref{fig:purity}) is approximately 70\% and mostly independent of the pileup, with the exception of the low \pt part of the distribution (below about 15~GeV), where the purity falls off as the pileup increases. As described above, the distribution is obtained using a set of very tight matching requirements between reconstructed tracks and simulated particles. However it is very likely for the track reconstruction algorithm to pick up at least one random hit, which actually will not change the track helix parameters in any significant way. If an alternative matching algorithm is used, based on comparing parameters of reconstructed tracks with those of simulated particles, then the purity increases to over 95\% in PU200 events.   

\section{Additional Studies}
In order to explore the limits of applicability of the proposed approach, event samples were generated with pileup up to twice the maximum expected for HL-LHC. The idea is to assess the robustness of the system by finding out where it breaks. It is important to mention that larger occupancy could be caused by different factors, not necessarily connected to a larger luminosity (e.g. larger than anticipated beam backgrounds, mismodeling in simulation, etc.).

Section (a) of Table~\ref{AllRoads} shows results of the simulation of the AM+FPGA approach applied to samples with increasing pileup multiplicity, including the expected value for the HL-LHC. It can easily be seen how the number of roads identified by the AM grows rapidly as the PU multiplicity is increased above the nominal value. At this point in time, we believe that it would be very challenging, if not impossible, to fit a track fitter engine in a single FPGA capable of dealing with more than 200 roads per LHC crossing. The limit is imposed by the maximum allowed latency. Therefore, most of these roads will have to be dropped with a consequent loss in efficiency. Section (b) of Table~\ref{AllRoads} shows the consequence of limiting the number of roads output by the AM to the first 200. It can be seen that, up to PU200, the efficiency is only very marginally affected, typically less that 1\%, but at PU300 we already see losses of 25\% and at PU400 the efficiency rapidly decreases. So, although at PU200 the system is performing well, there is not much headroom in case the detector occupancy turns out to be higher than expected. Below we describe a couple of potential solutions to address the problem. These solutions have been studied and shown to work in simulation. No attempts have been made to implement them in the demonstrator hardware, however at this time we do not see major obstacles that would make an implementation impossible. 

\begin{table}
\centering
\begin{tabular}{|c |c |c |c|}
\hline
Pileup & Number of roads   &  Tracking eff. for $\pt>10$\,GeV & Tracking eff. for $\pt>3$\,GeV \\ \hline
\multicolumn{4}{|c|}{(a) AM with all roads processed }\\ 
PU140 & 59 & 0.99 & 0.99    \\   
PU200 & 252  & 0.99 & 0.99  \\ 
PU300 & 12300  & 0.98 & 0.98  \\ 
PU400 & 60400  & 0.97 & 0.97   \\ \hline
\multicolumn{4}{|c|}{(b) AM with maximum of 200 roads }\\ 
PU140 & 56 & 0.99   & 0.99 \\ 
PU200 & 158 & 0.99  & 0.98 \\ 
PU300 & 200 &  0.87  & 0.73  \\ 
PU400 & 200  & 0.56   & 0.35   \\ \hline
\multicolumn{4}{|c|}{(c) AM+HT with maximum of 400 roads }\\ 
PU140  & 9 &    0.99   &   0.99     \\ 
PU200  & 13 &  0.99   &  0.99  \\ 
PU300  & 10  &    0.96  &  0.87      \\ 
PU400  & 8  &  0.73   &  0.54   \\ \hline
\multicolumn{4}{|c|}{(d) AM+$\delta s$ with maximum of 200 roads }\\ 
PU140  & 14  &    0.99   &   0.99     \\ 
PU200  & 40 &  0.99   &  0.99   \\ 
PU300  & 200  &    0.95  &  0.95      \\ 
PU400  & 200  &  0.95   &  0.90   \\ \hline
\end{tabular}
\caption{The average number of roads and the efficiency of reconstructing tracks with different \pt thresholds in four different scenarios: (a) all AM roads can be processed; (b) a maximum of 200 AM roads can be processed; (c) a maximum of 400 roads read out from the AM and filtered with the HT method can be processed; (d) a maximum of 200 AM roads with stub bend information can be processed. The results are shown for different pileup scenarios.}
\label{AllRoads}
\end{table}

\subsection{AM+Hough Transformation}
One of the keys to improving the performance of the tracking algorithm is to find a way to reduce the number of roads delivered to the track fitting stage while keeping the efficiency loss to an acceptable level. This is attempted by having the AM perform its pattern recognition function using only the transverse projection ($r$, $\phi$) of the stubs to find possible track candidates (roads) and then using a different method applied to the longitudinal projection ($r$, $z$), to confirm these candidates. As a second pattern recognition method we use a simplified version of the Hough Transform (HT)~\cite{HT}. The vast majority of the track candidates created by the AM is made of random combinations of stubs which align by chance in the transverse projection. Those will not be aligned, in general, in the longitudinal projection and will consequently be discarded even by a very simple pattern recognition algorithm. 

\begin{figure}
\centering
\includegraphics[width=16cm]{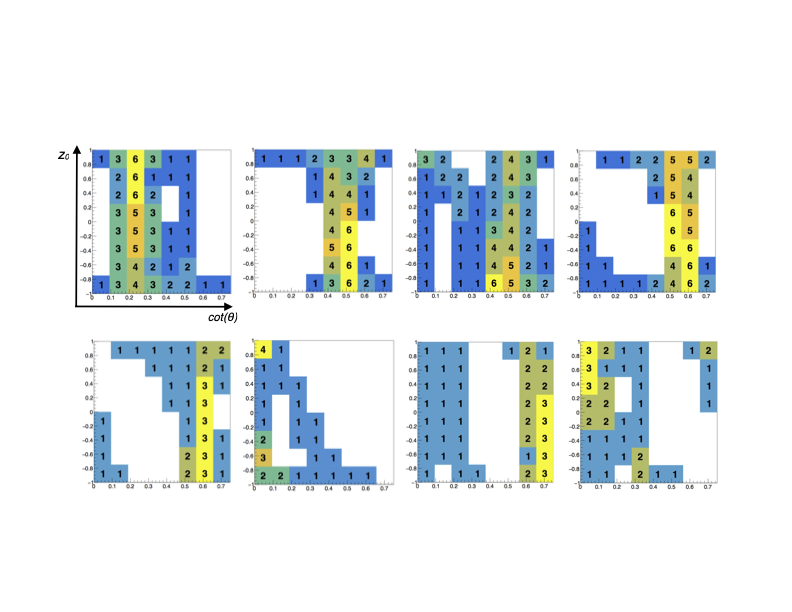}
\caption{Simplified Hough Transforms in ($cot(\theta), z_0$) space. Each square represents the Hough Transform applied to the stubs contained in a different road. $cot(\theta$) is the horizontal coordinate and $z_0$ is the vertical coordinate (re-scaled to the range of $-1.0\le z_0 \le 1.0$), as indicated by the axis labels on the top left plot. The numbers count how many detector layers contain at least one stub compatible with that particular region of track parameters (see text). We require at least five layers, out of six available, to accept that cell as a good track candidate. Top row: roads that were accepted and shipped to the track fitting engines because of at least one good track candidate. Bottom row: roads that were discarded because no good track candidate was found} 
\label{HoughTransform}
\end{figure}

The 2-dimensional longitudinal track parameter space ($cot(\theta), z_0$) in each trigger tower is subdivided into 64 rectangular cells, eight segments in $cot(\theta)$ and eight segments in $z_0$. The size of the individual segments is determined in order to obtain, on average, approximately the same number of stubs in each of them. 
Roads are processed one at a time. For every stub in a road, all the cells compatible with that stub are flagged. A cell is considered compatible with a stub if that stub could be generated by a track with parameters lying within that cell. There is a different flag for each cell and each detector layer. A good track will typically cause at least one cell to have all layers flagged, signaling the presence of a set of stubs, one per layer, compatible with a track from that particular cell in parameter space. In practice we need to store only a minimum and a maximum value of the $z$ coordinate of the stub for each cell and each detector layer. Those are obtained by detailed detector simulation. Figure~\ref{HoughTransform} shows a few roads where a track candidate was found and, therefore, the road was accepted and shipped to the track fitting engines (top row) and instances where a good track candidate was not found and the road was discarded (bottom row). Numbers count the detector layers which contain at least one stub compatible with belonging to a track with parameters within that particular cell. 
A cell is required to contain compatible stubs in at least five layers, out of six available, to be considered a good track candidate.

Section (c) of Table~\ref{AllRoads} shows what happens when doubling the maximum number of roads allowed from the AM but applying the HT filter at the same time. The number of roads delivered to the fitting stage is reduced by about one order of magnitude while the efficiency is increased by 10--20\%. A number of parameters (number of cells in the HT, maximum number of roads from the AM, etc.) could be further optimized. 

\subsection{Using Local Bend Information in AM}
In addition to the Hough Transform, an alternative concept for reducing the number of roads was studied in simulation. The approach relies on the idea of encoding additional information into the AM that was not previously utilized. Specifically, stub bend values tend to be small for high \pt tracks and vice versa. For the real tracks, the values of $\delta s$  measured in different detector layers have to be consistent with each other. Fake roads, on the other hand, typically contain stubs generated by more than one particle and thus having inconsistent stub bends across different layers. 

In order to include bend information into the pattern, we divide the possible range of $\delta s$ values into three equal width bins for the two inner layers of the tracker, five bins for the two middle layers, and seven bins for the two outer layers. This binning scheme yields the best results in terms of efficiency and road multiplicity. The bend information is then included into the AM pattern superstrip definition (along with the spatial coordinates) and a new pattern bank is generated. Section (d) of Table~\ref{AllRoads} shows the results. The stub-bend-based method yields similar results to the AM+HT approach at PU300 and is superior at PU400. It should be noted, however, that one has to be careful when using $\delta s$ as this may introduce additional dependency on the modeling of stub formation.

In summary, both AM+HT and AM+$\delta s$ studies illustrate that one could get this system to perform satisfactorily up to PU400, that is, a 100\% increase in the maximum occupancy can be tolerated with respect to the expected HL-LHC environment.

%% file: Summary.tex
\section{Summary \label{sec:summary}}
We presented an end-to-end solution that allows the inclusion of tracking information at the L1 trigger for HL-LHC, namely reconstructing particle tracks using data from a silicon tracker at an input event rate of 40\,MHz and with a latency of less than 4 microseconds. The solution relies on Associative Memories to implement a pattern recognition algorithm that quickly identifies stubs in the detector that are associated with real tracks. A number of novel algorithmic approaches have been presented in order to maximize the performance and to reduce the overall size of the system. These novel concepts prescribe better ways to partition the detector into trigger towers, to define Associate Memory patterns, and to perform track fitting with a small set of constants derived from a principal component analysis. 

The solution has been tested successfully through a hardware demonstration system featuring many innovations. The ATCA-based demonstration system relies on a full-mesh ATCA architecture in order to efficiently partition and dispatch the data for regional and time multiplexing. It utilizes newly supported read/write features in Xilinx block RAMs to provide for low latency stub storage and retrieval, features a fully pipelined and cycle-accurate emulation of AM operations implemented in FPGA, and includes ultra-fast track fitting engines implemented using cascading DSP operations. 

All the presented performance metrics for the system were found to be excellent and well within the tight specifications needed for the successful running of such a trigger at the HL-LHC. Additional simulation based studies presented in this paper outline a path for future improvements. The described solution is highly scalable and thus has a potential to address needs of the future high energy collider physics experiments beyond the HL-LHC.

\section{Acknowledgements \label{sec:acknowledgements}}
We are grateful to the CMS Collaboration for use of their detector simulation software. 

%% file: Hardware.tex
\section{Demonstration Hardware\label{sec:hardware}}

\begin{figure}
\centering
\includegraphics[width=10cm]{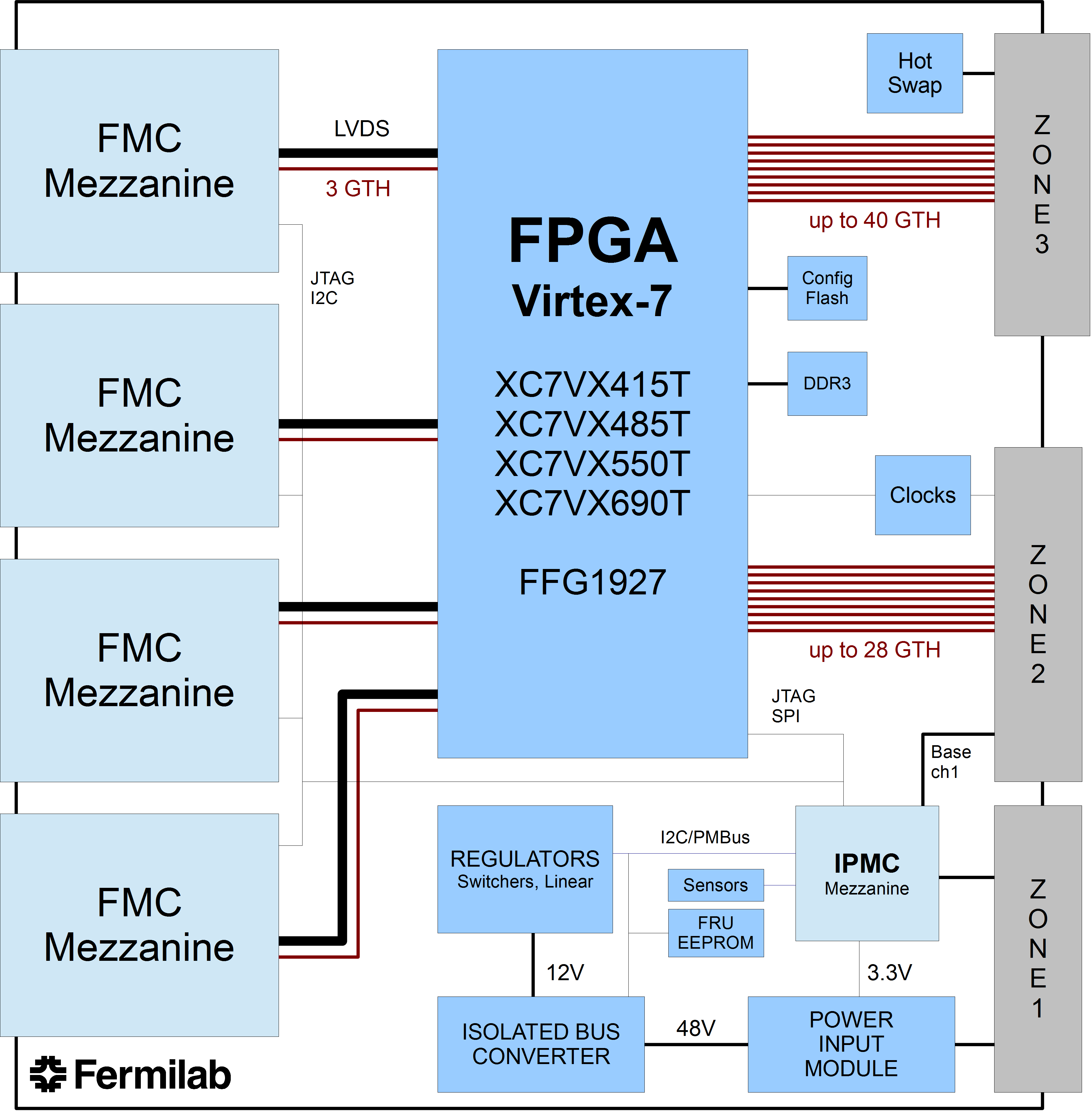}
\caption{The Pulsar IIb front board block diagram. GTH indicates a particular type of the high speed serial transceivers featured in Xilinx FPGAs.}
\label{pulsar2b_block}
\end{figure}

The Pulsar IIb board is a custom ATCA full-mesh enabled FPGA-based processor board which has been designed with the goal of creating a scalable architecture abundant in flexible, non-blocking, high bandwidth interconnections. Initially motivated by silicon-based tracking trigger needs for the LHC experiments, the Pulsar IIb board is uniquely positioned as a flexible R\&D platform ideally suited to situations where multiple processing engines need to be tightly coupled. The full-mesh backplane provides a very fast and efficient connection between the Pulsar IIb boards and opens up options for data sharing in both space and time. The Pulsar IIb board block diagram is shown in Figure~\ref{pulsar2b_block}.

A single large Xilinx Virtex-7 FPGA forms the heart of the Pulsar IIb board. The FPGA used on the board (\texttt{XC7VX690T-2FFG1927C}) features 693k logic cells, 3600 DSP slices, 52\,Mb/s of dual port BlockRAM, and 80 multi-gigabit transceivers (MGTs) which support line rates up to 11.3\,Gb/s. The worst case power dissipation in the FPGA is on the order of 60\,W with approximately 40\,W for the transceivers and 20\,W for the regular I/O and core logic. 

ATCA boards are designed for high availability operation. This is achieved by using a well-defined and robust system for control and monitoring, called the intelligent platform management interface (IPMI). All modules in the ATCA shelf must communicate with the shelf manager board using the IPMI protocol. This protocol is used, among other things, to coordinate hot-swap insertion and removal of boards, report sensor readings, define alarm thresholds, and actively control fan speeds. All ATCA boards must include a microcontroller to support the IPMI protocol. The Pulsar IIb board provides a connector for a small intelligent platform mezzanine card (IPMC) for this purpose. The IPMC used on the Pulsar IIb is a custom design developed at Fermilab. The form factor and pinout of the IPMC mezzanine matches similar cards developed by CERN and other organizations. The Fermilab IPMC uses an ARM Cortex M3 microcontroller and includes a fast Ethernet PHY chip for network connectivity and a microSD flash card for bulk storage.

On the Pulsar IIb board, 28 MGT transceivers are routed directly from the FPGA to the full-mesh fabric interface backplane channels. Most 14-slot ATCA full-mesh backplanes provide four bidirectional serial ports (lanes), however on the Pulsar IIb board only two of these backplane channels are utilized for slot to slot communication. The direct connection between the FPGA and the full-mesh backplane enables the use of the built-in MGT diagnostics to evaluate the quality of the full-mesh links at speeds up to 10\,Gb/s.  Communication between Pulsar IIb boards has been sucessfully tested across the entire width of the ATCA backplane with line rates up to 10\,Gb/s.

The Pulsar IIb board supports up to four FPGA Mezzanine Cards (FMC). Mezzanine cards may contain FPGAs, ASICs, optical transceivers, or any other custom hardware.  Each FMC mezzanine card connector is wired directly to the main FPGA on the Pulsar IIb board and supports 36 user-defined LVDS pairs and three 10\,Gb/s serial bidirectional lanes. Mezzanine cards are connected to the Pulsar IIb board JTAG bus and support local and remote programming over the network through the IPMC. A general purpose I2C bus connects each mezzanine card to the main FPGA on the Pulsar IIb board. The PRM card (Fig.~\ref{pulsar2}) contains two Xilinx Ultrascale Kintex KU060 FPGAs and a socket that allows for the possible future integration of VIPRAM prototypes~\cite{Hoff:2017vib}. 

The Rear Transition Module (RTM) hosts ten Quad Small Form Factor (QSFP+) pluggable transceivers. These optical (or passive copper) modules connect directly to the Pulsar IIb board FPGA MGT transceivers and support bidirectional data rates of up to 400\,Gb/s. The RTM is compliant with the PICMG 3.8 RTM specification and is considered an intelligent field replaceable unit (FRU) supporting hot swap insertion and removal. A small ARM Cortex M3 microcontroller is used on the RTM to monitor QSFP+ modules, voltage regulators, and temperature sensors. The RTM microcontroller communicates with the Pulsar IIb board IPMC module.

